\documentclass{jfm}
\usepackage[dvipsnames]{xcolor}
\usepackage{color}
\usepackage[utf8]{inputenc}
\usepackage{amsmath}
\usepackage{amssymb}
\usepackage{authblk}
\usepackage{float}
\usepackage{graphicx}
\usepackage{enumerate}
\usepackage[colorinlistoftodos]{todonotes}
\usepackage{appendix}
\usepackage{upgreek}
\usepackage{color, colortbl}
\usepackage{subcaption}
\usepackage{etoolbox}

\floatstyle{plain}
\restylefloat{figure}

\newcommand{\prm}[1]{#1}
\newcommand{\var}[1]{{#1}}

\definecolor{Gray}{gray}{0.9}
\definecolor{my_emerald}{RGB}{1,99,52}

\usepackage[inline]{trackchanges}
\addeditor{LK}
\addeditor{RA}
\addeditor{LC}

\title{On efficient asymptotic modelling of thin films on thermally conductive substrates}

\author[1]{Ryan H. Allaire}
\author[1]{Linda J. Cummings}
\author[1]{Lou Kondic}
\affil[1]{\textit{New Jersey Institute of Technology, Department of Mathematical Sciences and Center for Applied Mathematics and Statistics, Newark NJ, 07102}}
\date{}
\begin{document}

\maketitle

\begin{abstract}
We consider a free surface thin film placed on a thermally conductive substrate and exposed to an external heat source in a setup where the heat absorption depends
on the local film thickness.  Our focus is on modeling film evolution while
the film is molten.  
The evolution of the film modifies local heat flow, 
which in turn may influence the film surface evolution through thermal variation of the 
film's material properties.  Thermal conductivity of the substrate plays 
an important role in determining the heat flow and the temperature field
in the evolving film and in the substrate itself.  In order to reach a tractable formulation, 
we use asymptotic analysis to develop a novel thermal model that 
is accurate, computationally efficient, and that accounts for the heat flow in both the
in-plane and out-of plane directions.  We apply this model to metal 
films of nanoscale thickness exposed to heating and melting by laser pulses, 
a setup commonly used for self and directed assembly of various metal 
geometries via dewetting while the films are in the liquid phase. 
We find that thermal effects play an important role, and in particular that the inclusion of 
temperature dependence in the metal viscosity modifies the time scale of the evolution 
significantly.  On the other hand, in the considered setup the Marangoni (thermocapillary) effect 
turns out to be insignificant. 
\end{abstract}

\section{Introduction}\label{intro_section}
 The dynamics of thin liquid films is a topic of extensive interest with a number of applications ranging from biomedical \citep{Li2016,Chen2016} to electronic coatings and nanotechnology \citep{Zhang2010}. The inclusion of thermal effects in thin film dynamics, relevant for many applications, is a mathematically challenging problem. To develop a realistic model one must consider multiple factors, such as the heat supply mechanism(s), possible dependence of material parameters on temperature, heat loss mechanisms, and phase changes. When the liquid of interest is placed upon a thermally conductive substrate one must also account for the heat flow within the substrate as well as the interaction between the liquid and the substrate. Numerous models have been developed to address these complications using continuum theory, which in general describes both the thermodynamics and fluid dynamics in terms of partial differential equations (PDEs), derived from first principles. In situations where there is a small aspect ratio (ratio of typical film thickness to typical lateral length scale of interest) long-wave theory (LWT) may be used, which effectively enables the fluid dynamics problem to be reduced to a 4th order PDE for film thickness. LWT has already proved very valuable in a variety of settings such as liquid crystals, paint coatings, tear-films, nanotechnology and many others (see \citet{craster_rmp09} for a comprehensive review). Due to the variety of length and time scales present, the applicability of LWT to the problem of heat conduction in a thin liquid film is not always clear-cut. Of the issues outlined above we highlight the following in this work: (i) the influence of temperature on film evolution; (ii) heating/cooling mechanisms; and (iii) the application of LWT to heat conduction. 

Various thermal effects that may influence the evolution of the film thickness have been considered in prior work. For an isothermal nanoscale film the primary dewetting mechanism is liquid-solid interaction, often modeled by a disjoining pressure (see~\cite{Israelachvili} for an extended review). For non-isothermal films, gradients in temperature may give rise to surface tension gradients (thermocapillary or Marangoni effects), which develop when heating from below \citep{scriven_sternling_1964} and can destabilize the film. The work of \citet{shklyaev12} finds novel stability thresholds between monotonic and oscillatory instabilities (in both cases, in the linear regime the instability grows as $e^{\omega t}$ with $\Re(\omega)>0$, but $\Im (\omega)$ is zero in the former case and nonzero in the latter) that also account for heat losses from the free surface of the film (referred to here as radiative heat losses). In that work, the film is heated from below via a constant heat flux from a substrate of much lower thermal conductivity.
\citet{batson2019} perform a stability analysis similar to that of \citet{shklyaev12}, but model the substrate explicitly rather than as a simple boundary condition. They solve a full heat equation for the substrate temperature, and find that oscillatory instabilities arise primarily due to thermal coupling between the film and the substrate. A number of other works have considered the coupling between the evolution of film and substrate temperatures.  \citet{Saeki2011}, for example, consider a film/substrate system heated by a laser and find that the rate of change of film reflectivity $R$ with thickness $h$, $\mathrm{d}R/\mathrm{d}h$, may promote either stability or instability of the film depending on the sign of $\mathrm{d}R/\mathrm{d}h$. The magnitude of the incident laser energy was earlier shown to influence film thickness evolution by \citet{Oron2000}, who showed in particular that increasing the laser energy can partially inhibit film instability.

Another important effect that may influence film evolution is the dependence of material parameters, such as density, thermal conductivity, surface tension, heat capacity and viscosity, on temperature. These relationships are often assumed to be linear, although a strongly nonlinear Arrhenius-type dependence of viscosity on temperature may exist. \citet{oron_rmp97} formulated a thin film model in which viscosity variation is included, and in later work \citet{Seric_pof2018} found that film evolution is strongly affected by the inclusion of temperature-dependent viscosity. If temperature variations are sufficiently large, the film may undergo a phase change (liquefaction or solidification). This has been considered using a variety of approaches, for example \citet{trice_prb07} use a latent heat model to describe such phase change whereas others, such as \citet{Seric_pof2018}, assume phase change to be instantaneous. 
 
Modelling of heat losses in a liquid film often focuses on the boundary effects, since viscous dissipation can usually be ignored.
Radiative heat losses from the liquid to the surrounding medium are typically modelled by a Robin type boundary condition \citep{Saeki2011, Saeki2013, Oron2000, shklyaev12, atena09} whereas the heat loss/gain from the film to the substrate has been modelled variously by (i) a constant temperature \citep{oron1998, Oron2000, Saeki2011}, (ii) constant flux \citep{atena09, shklyaev12}, or (iii) continuity of temperatures and fluxes, known as perfect thermal contact \citep{trice_prb07,Seric_pof2018, Dong_prf16,Saeki2011}. The choice of boundary conditions plays an important role when formulating and solving equations to describe the heat flow. 

In many cases an asymptotic approach may be adopted, giving rise to simplified leading order temperature equation(s). The work of \citet{Saeki2011}, for example, includes both radiative heat losses and heat transfer at the liquid-solid interface, and gives rise to a depth-averaged ($z$-direction) equation for film temperature, which retains parametric $z$-dependence even when radiative heat losses are ignored. In later work \citet{Saeki2013} developed similar leading order equations for film temperature when the film is optically transparent. In this case the film temperature dependence on $z$ is slaved to the inclusion of radiative heat losses. \citet{trice_prb07}, on the other hand, conclude that using a $z$-independent film temperature model is sufficient when radiative heat losses can be neglected and film-to-substrate heat losses are dominant (e.g. when there is a high thermal conductivity ratio between the film and substrate). These previous works demonstrate that boundary conditions play an integral role in the asymptotic formulation of a model and may facilitate simple models that eliminate $z$-dependence (e.g. \citet{shklyaev12}).

 Due to the small aspect ratio of the film, a commonly used ``reduced" model for heat conduction is one that neglects in-plane diffusion altogether \citep{trice_prb07, Dong_prf16, Seric_pof2018}. This model, which we refer to here as (1D), is much simpler than a model that includes full heat diffusion and is typically justified by arguing that in-plane diffusion occurs on a much longer time scale than that of out-of-plane diffusion. Alternative simplified models have also been proposed. The work of \citet{shklyaev12}, for example, uses LWT to derive evolution equations for heat conduction that differ significantly from the 1D model of \citet{Dong_prf16} (and from the asymptotic model considered in this paper).  \citet{atena09} also derive leading order temperature equations that do not rely on the 1D approximation. More recent work by \citet{Seric_pof2018} briefly compares predictions from (1D) with those using a full thermal diffusion model, and suggests that (1D) performs poorly by comparison, though the analysis is far from complete. Despite the extensive literature, the scenarios for which thermal diffusion model (1D) is valid remain unclear. {\it A key objective of the present paper is to present a thermal model for thin film flow that includes in-plane heat conduction at reasonable computational complexity} and to compare with both (1D) and with the full heat diffusion model (which serves as a benchmark).
  
  In this paper, we consider films placed upon a thermally conductive substrate and heated by a laser. Heat generation by a laser source is complicated to model and requires in general that one accounts for the optical properties of the film, such as reflectivity, transmittance, and absorption. These properties may depend on refractive indices of the air, film and substrate, as well as the respective extinction coefficients. Again, various modelling approaches have been taken in the literature: we note for example that \citet{Saeki2011, Saeki2013} present a detailed model for laser energy that involves complicated expressions for the optical properties; whereas \citet{trice_prb07} propose a simpler approach (to be discussed later) in which these properties are approximated. An important application of laser heating is pulsed laser induced dewetting (PLiD) of metal films. The mechanism by which liquid metals evolve into assemblies of droplets has been explored via experiments~\citep{henley_prb05}, simulations \citep{Seric_pof2018, Dong_prf16}, and theory~\citep{trice_prb07} with applications ranging from nanowire growth \citep{kim_nanolet2009,ross_iop_2010, Shirato2011}, to plasmonics \citep{halas_chemrev2011} and photovoltaics \citep{atwater_natmat10}; see also \citet{Hughes2017} and~\citet{Makarov2016} for recent application-centered reviews, and \citet{ruffino_nano19} and~\citet{kondic_arfm_2020} for reviews focusing 
  on molten metal film instabilities. Of late, PLiD has been used to organize nanoparticles into patterns of droplets via Rayleigh-Plateau type instabilities~\citep{favazza_nanotech06,mckeown12, ruffino_matlet2012}, induced by exposing metal films/filaments upon (typically) Si/SiO$_2$ substrates to laser irradiation, effectively liquefying the film for tens of nanoseconds. The liquefied film breaks up into droplet patterns, which then resolidify, freezing the patterns in place. Thermal effects are found to be highly relevant, influencing the stability, evolution, and final (solidified) configurations of molten metal films (see for example, \citet{trice_prb07}). A number of experimental studies have considered metallic systems such as Co \citep{favazza_nanotech06,favazza_apl06, favazza_JEM06,trice_prb07}, Ag \citep{krishna_nanotech10}, Au \citep{Yadavali2013}, Ni \citep{Nanoscale12}, as well as multi-metal systems \citep{fowlkes_10}. The large variety of experimental work that has been done on nanoscale metal films calls for a firm theoretical foundation, which can both explain existing results and suggest new approaches. 
 
The focus of the present paper is development of a consistent, asymptotically valid, mathematical model that accounts for (i) heat absorption that is influenced by the local value of (time-dependent) film thickness; (ii) in-plane and out-of-plane heat diffusion in a tractable manner; (iii) self-consistent coupling of the heat flow and film evolution, and (iv) thermal variation of material properties, in particular of surface tension and viscosity. Long-wave theory (LWT) is used to reduce modelling of the film evolution to a 4th order PDE for the film thickness and to develop an asymptotic model for heat conduction. We consider a setup where the primary heat loss mechanism is through the substrate rather than the liquid-air interface, and the thermal conductivity of the film is much higher than of the substrate, as appropriate for metal films on SiO$_2$ substrates. We will show that the proposed model (called asymptotic model (A) in what follows) produces accurate results, while avoiding the shortcomings of models that ignore coupling of fluid dynamics and thermal transport and producing results with a reasonable computational effort. It should be emphasized that the use of a more complex model (called full (F) model below) is orders of magnitude more computationally expensive (even for small computational domains, the computing time is measured in days on a modern computational workstation (in a serial mode)). Our asymptotic model provides essentially indistinguishable results at a fraction of the computational cost. 

The rest of the paper is organized as follows. 
In \S\ref{model_formulation_section} we formulate a general mathematical model by introducing appropriate scales, the corresponding dimensionless system, and relevant dimensionless parameter groups. We present three different models of heat conduction: a full diffusion model (F), a 1D diffusion model (1D), and an asymptotic model (A); and we summarize the derivation of the thin film evolution equation (the fluid mechanical model always used), accomplished using LWT and accounting for the possibility of temperature dependence of material parameters. Section \ref{results_section} contains our main results. In \S\ref{lsa_section} we perform linear stability analysis (LSA) on the film evolution equation to understand the circumstances under which disturbances to the liquid film lead to instabilities, and to predict the manner of film breakup. In \S\ref{simulation_setup_section} we summarize the conditions under which our simulations are carried out, and in \S\ref{Model Comparison Section} we display results comparing the three models for heat conduction. In \S\ref{Experimental_comparison_section} we restrict attention to the asymptotic model for heat conduction and study how temperature dependence of both viscosity and surface tension influence the results. We find that temperature dependence of the viscosity has the most significant  effect on the instability development, while temperature-induced variation of surface tension plays only a minor role. Furthermore, in the physically-relevant regime, allowing viscosity to vary with temperature produces films that dewet fully in the liquid phase, while if viscosity is fixed at its melting temperature value the dewetting occurs much closer to the solidification time, which may result in only partial drop formation. We conclude in \S\ref{conclusions} with a brief summary and discussion.

\section{Model Formulation}\label{model_formulation_section}
\begin{figure}[H]
    \centering
    \includegraphics[width=0.6\textwidth]{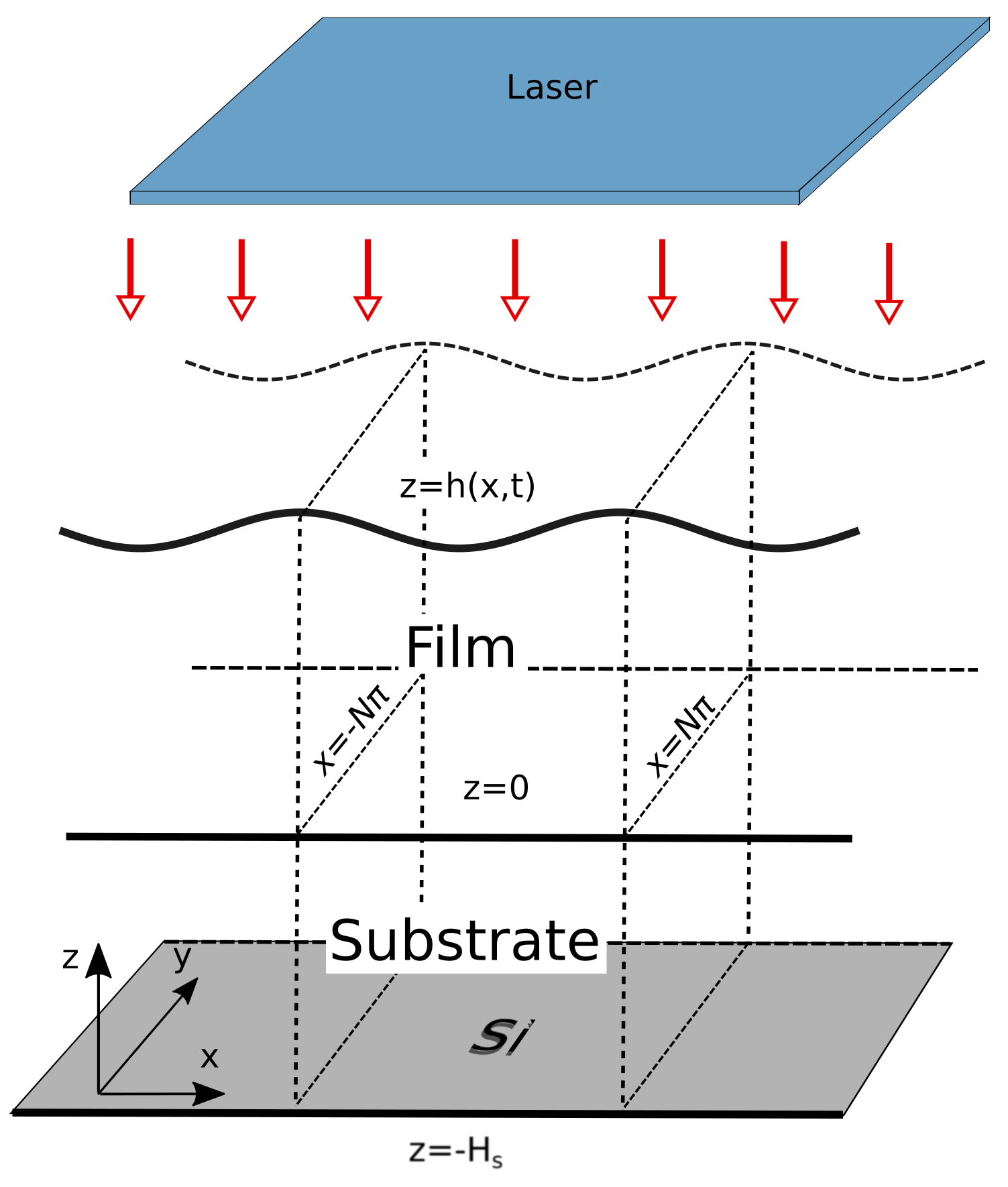}
    \caption{3D schematic of the film, substrate and laser system. In dimensionless units the mean film thickness is equal to $1$, the substrate thickness is given by $H_{\rm s}$ and the domain width is $2N\pi$ (both $N=1$ and $N=20$ will be used in simulations). The model is presented in 3D but for simplicity simulations will be performed only in 2D.}
    \label{fig:schematic}
\end{figure}

Consider a molten metal film (assumed initially solid) of characteristic lateral length-scale $\prm{L}$, and (nanoscale) thickness $\prm{H}$, heated by a laser, and in contact with a thermally conductive solid substrate of finite thickness, which itself rests upon a much thicker Si slab. The basic setup is sketched in Figure~\ref{fig:schematic}. Here we consider the substrate to be thin, and comparable in size to the film thickness, $H$. We define the aspect ratio of the film to be $\epsilon=H/L \ll 1$.

 In the following we refer to the in-plane coordinates as $x,y$ and the out-of-plane coordinate as $z$. For completeness, we present the governing equations for a 3D system, though the results presented in this paper will be for the 2D case in which all quantities are independent of $y$. We define $L$, $H$, $U$, $\epsilon U$, $t_{\rm s}$, $T_{\rm melt}$, $\mu_{\rm f} U/(\epsilon^2 L)$ and $\gamma_{\rm f}$ (where $\mu_{\rm f}$ and $\gamma_{\rm f}$ are the viscosity and surface tension of the film at melting temperature, $T_{\rm melt}$) to be the in-plane length scale, out-of-plane length scale, in-plane velocity scale, out-of plane velocity scale, time scale, temperature scale, pressure scale, and surface tension scale, respectively (the values of the material parameters used are given in Table \ref{table:ref_paras} in \S\ref{appendix_scales}). Similar to \citet{lang13}, we set $t_{\rm s}=3 \mu_{\rm f} L/ (\epsilon^3 \gamma_{\rm f})$, which can be interpreted as a viscous time scale. The in-plane velocity scale is fixed as $U=\epsilon^3 \gamma_{\rm f}/(3 \mu_{\rm f})$ so that $t_s=L/U$.
  The length scale is fixed as $L=\lambda_{\rm m}/(2\pi)$, where $\lambda_{\rm m}$ is the most unstable wavelength obtained from linear stability analysis (LSA) with surface tension and viscosity fixed as $\gamma_{\rm f}$ and $\mu_{\rm f}$, respectively (see \S\ref{appendix_lsa} for details). We treat the film as an incompressible Newtonian fluid and assume that viscosity is independent of $z$. The resultant dimensionless system then comprises the following fluid equations, which hold on $0<z<h$,
 \begin{align}
    \epsilon^2 {\mathrm{Re}} \left( \partial_t u + \boldsymbol{u} \bcdot \bnabla u \right) &= -\partial_x p + \epsilon^2 \nabla_2 \cdot \left( \mathcal{M} \nabla_2 u \right) + \mathcal{M}  \partial_z^2 u + \epsilon^2 \partial_y \mathcal{M} \partial_x v, \label{system1} \\
    \epsilon^2 {\mathrm{Re}} \left( \partial_t v + \boldsymbol{u} \bcdot \bnabla v \right) &= -\partial_y p + \epsilon^2 \nabla_2 \cdot \left( \mathcal{M} \nabla_2 v \right) + \mathcal{M} \partial_z^2 v + \epsilon^2 \partial_x \mathcal{M} \partial_y u, \label{system2} \\
    \epsilon^4 {\mathrm{Re}} \left( \partial_t w + \boldsymbol{u} \bcdot \bnabla w \right) &= -\partial_z p +  \epsilon^4 \nabla_2 \cdot \left( \mathcal{M} \nabla_2 w \right) + \epsilon^2 \mathcal{M} \partial_z^2 w + \epsilon^2 \nabla_2 \mathcal{M}\cdot \partial_z(u,v), \label{system3} \\
    \bnabla \bcdot \boldsymbol{u} &= 0, \label{system4}
\end{align}
the following equations of heat conduction,
\begin{align}
        \epsilon^2 \mbox{Pe}_{\rm f} \partial_t T_{\rm f} &= \epsilon^2 \nabla_2^2 T_{\rm f} + \partial_z^2 T_{\rm f} + \epsilon^2 Q, \quad &&\mbox{for\ }\quad z \in \left( 0,h \right), \label{system5} \\
     \mbox{Pe}_{\rm s}\partial_t T_{\rm s} &= \epsilon^2 \nabla_2^2 T_{\rm s} + \partial_z^2 T_{\rm s}, \label{system6} \quad &&\mbox{for\ }\quad  z \in \left( -H_{\rm s},0 \right),
\end{align}
and boundary conditions,
\begin{align}
       w&=\partial_t h + u\partial_x h + v \partial_y h,    \quad &&\mbox{on\ }\quad  z=h, \label{system7} \\
    \left[ \boldsymbol{n} \cdot \mathsfbi{\var{T}} \cdot \boldsymbol{n} \right] &= 3\gamma \left( \bnabla \bcdot \boldsymbol{n} \right) - 3\Pi(h), \quad &&\mbox{on\ }\quad z=h,
    \label{system8} \\
        \epsilon \left[ \boldsymbol{n} \cdot \mathsfbi{\var{T}} \cdot \boldsymbol{t}_1 \right] &= -3 \partial_x \var{\gamma}, \label{system9_1} \quad &&\mbox{on\ }\quad z=h, \\
            \epsilon \left[ \boldsymbol{n} \bcdot \mathsfbi{\var{T}} \bcdot \boldsymbol{t}_2 \right] &= - 3\partial_y \var{\gamma}, \label{system9_2} \quad &&\mbox{on\ }\quad z=h, \\
    \boldsymbol{n} \bcdot \bnabla T_{\rm f} &= 0, \quad &&\mbox{on\ }\quad z=h, \label{system10} \\
    \partial_z T_{\rm f} &= \mathcal{K} \epsilon^2 \partial_z T_{\rm s}, \quad T_{\rm f} = T_{\rm s}, \quad &&\mbox{on\ }\quad z=0, \label{system11} \\
     \boldsymbol{u} &= \boldsymbol{0}, \quad &&\mbox{on\ }\quad z=0, \label{system12} \\
    \partial_z T_{\rm s} &= {\rm Bi} \left( T_{\rm s} - \prm{T}_{\rm a} \right), \quad &&\mbox{on\ }\quad z=-H_{\rm s}, \label{system13} \\
    \partial_x T_{\rm f} &= \partial_x T_{\rm s}=0, \quad &&\mbox{on\ }\quad x= \pm N \pi, \label{system14} \\
    \partial_y T_{\rm f} &= \partial_y T_{\rm s}=0, \quad &&\mbox{on\ }\quad y=\pm N \pi. \label{system15}
 \end{align}
 Here, the fluid velocity is given by $\boldsymbol{u}=(u,v,w)$, pressure by $p$, and film and substrate temperatures by $T_{\rm f}$ and $T_{\rm s}$, respectively. Subscripts ${\rm f}$ and ${\rm s}$ stand for film and substrate, respectively, unless otherwise stated and $\boldsymbol{0}=(0,0,0)$. We refer to the gradient operator as $\bnabla=(\partial_x, \partial_y, \partial_z)$, its in-plane counterpart as $\bnabla_2=(\partial_x,\partial_y,0)$, and the in-plane Laplacian operator as $\nabla_2^2$ (defined by $\nabla_2^2 u = \partial_x^2 u + \partial_y^2 u$ for a given scalar function $u$). Equations \eqref{system1}--\eqref{system6} are the  Navier-Stokes (NS) equations representing conservation of mass and momentum for the film, together with thermal energy conservation in both film ($0<z<h$) and substrate ($-H_{\rm s}<z<0$) domains, both of lateral extent
  $2N\pi$, $-N\pi<x,y<N\pi$ (for simulations we use either $N=1$ or $N=20$, but $N$ can be any positive integer). The unit vector $\boldsymbol{n}$ denotes the outward normal to the film's free surface, $z=h$. The equations above introduce the following dimensionless parameters:
  
 \begin{align}
         {\rm Re} &= \frac{\prm{\rho}_{\rm f}\prm{U} \prm{L}}{\prm{\mu}_{\rm f}}, \qquad \mathcal{M} = \frac{\mu}{\mu_{\rm f}}, \qquad \mathcal{K}=\frac{\prm{k}_{\rm s}}{\prm{k}_{\rm f}}\epsilon^{-2}, \label{nd_parameters} \\ \mbox{Pe}_{\rm f} &= \frac{\left( \prm{\rho} \prm{c} \right)_{\rm f} U L }{\prm{k}_{\rm f} }, \qquad \mbox{Pe}_{\rm s} = \frac{\left( \prm{\rho} \prm{c} \right)_{\rm s} U \epsilon H }{\prm{k}_{\rm s} }, \qquad \mbox{Bi} = \frac{\prm{\alpha} \prm{H}}{\prm{k}_{\rm s}},  \nonumber 
 \end{align}
 which are the Reynolds number, dimensionless viscosity, (scaled) thermal conductivity ratio, Peclet numbers, and Biot number, respectively. We assume $\epsilon^2 {\rm Re} \ll 1$; the remaining quantities in Eq.~\eqref{nd_parameters} are assumed $O(1)$. For further discussion of the choice of scales and parameter values see \S\ref{appendix_scales}, in particular Table~\ref{table:ref_paras}, and \S\ref{asymptotic_model_section} later. The definitions of the laser source term $Q$ and the disjoining pressure $\Pi(h)$ are given in the discussion below. The material parameters $(\rho,c,k)_{{\rm f},{\rm s}}$ represent the density, specific heat capacity, and thermal conductivity of the film and substrate, respectively. We assume that the substrate is optically transparent and does not absorb laser energy.

 In experiments a Si substrate is often used \citep{wu_lang11,henley_prb05,Yadavali2013} on top of which a native layer of oxide, usually SiO$_2$, $3$-$4$ nm in thickness, typically exists (though an additional oxide layer, typically about $100$ nm thick, may also be deposited). Below the oxide in either case is Si, which has a much higher thermal conductivity, and can therefore be assumed isothermal relative to the SiO$_2$. Consistently, we consider the (SiO$_2$) substrate to be positioned on top of a thick 
layer of much higher conductivity, assumed to be at constant ambient temperature, $T_{\rm a}$. We model the heat loss from the top (SiO$_2$) substrate to the thick Si layer below via a Newton law of cooling at $z=-H_{\rm s}$ (Eq.~\eqref{system13}) with Biot number, ${\rm Bi}$ (related to the dimensional heat transfer coefficient, $\alpha$). The value of ${\rm Bi}$ was chosen so that the film melts and solidifies on a time scale comparable to the film evolution (although the value ${\rm Bi}=2\times 10^{-3}$ is presented, the range $7 \times 10^{-4}$-$7\times 10^{-3}$ was considered). We assume the following form of the heat source, $Q$ in Eq.~\eqref{system5}, representing the external volumetric heating due to the laser at normal incidence
(see \citet{trice_prb07,Seric_pof2018}),
 \begin{align}
     Q &= F(t) \left[ 1-R(h) \right] \exp{\left[-\alpha_{\rm f}\left(h-z \right)\right]}, \label{Q_eqn} \\
    F(t) &= C \exp \left[ -\left(t-t_{\rm p} \right)^2/(2\sigma^2) \right], \nonumber \\
    C &= \frac{E_0 \alpha_{\rm f} L^2}{ \sqrt{2 \pi}\sigma t_{\rm s} H k_{\rm f}   T_{\rm melt}}, \nonumber
 \end{align}
where $C$ is a constant (assumed $O(1)$) proportional to the amount of incident energy, $E_0$, applied from the laser onto the film, $\alpha_{\rm f}^{-1}$ is the (scaled) absorption length for laser radiation in the film, and $F(t)$ captures the temporal power variation of the laser, taken to be a Gaussian pulse centred at $t_{\rm p}$ and of width defined by $\sigma=t_{\rm p}/ ( 2 \sqrt{2 \ln{2}} )$.  Similar to prior work by a number of authors \citep{Saeki2011,Saeki2013,trice_prb07,Dong_prf16,Seric_pof2018, Oron2000} the transmittance of laser-source heating is modeled via the Bouguer-Beer-Lambert Law (see, {\it e.g.} \citet{howell2010}), which in Eq.~\eqref{Q_eqn} is presented as a spatially-dependent source term, $\exp(-\alpha_{\rm f}(h-z))$. In general the reflectivity of the film, $R(h)$, on a transparent substrate, can be determined by solving Maxwell's equations with the appropriate boundary conditions \citep{heavens55}. The resultant form is quite cumbersome to work with, however, 
and instead we approximate $R(h)$ by the simple functional form \citep{Seric_pof2018}
\begin{align}
 R(h)=r_0 \left(1-\exp \left(-\alpha_{\rm r} \var{h} \right) \right), \nonumber
\end{align}
 where $r_0$ and $\alpha_{\rm r}$ are dimensionless fitting parameters, determined by a least-squares fit of the approximate $R(h)$ to the full expression for reflectivity.  

Equations \eqref{system7}--\eqref{system10} are boundary conditions on the free surface, $z=h(x,y,t)$, with unit normal $\boldsymbol{n}=\bnabla \left( z-\epsilon h \right)/|\bnabla \left(z- \epsilon h\right)|$ and tangent vectors $\boldsymbol{t}_1$ and $\boldsymbol{t}_2$ given by $\boldsymbol{t}_1 = (1,0,\epsilon \partial_x h)/\sqrt{1+\epsilon^2 (\partial_x h)^2}$ and $\boldsymbol{t}_2 =(0,1,\epsilon \partial_y h)/\sqrt{1+\epsilon^2 (\partial_y h)^2}$. The kinematic boundary condition (KBC) is given by Eq.~\eqref{system7}; Eqs.~\eqref{system8}, \eqref{system9_1} and \eqref{system9_2} are the dynamic boundary conditions, representing a balance of stress between the liquid and air phases, where $\mathsfbi{T}$ is the Newtonian stress tensor, $\gamma$ is the surface tension, and $\Pi(h)$ is the disjoining pressure representing liquid-solid interaction. Many forms of $\var{\Pi}(\var{h})$ are used in the literature\footnote{For more information regarding the microscopic nature of the disjoining pressure, 
we refer the reader to~\citet{Israelachvili}.}; we take 
 \begin{align}
     \var{\Pi}(\var{h})= {\rm K} \left[ \left( \frac{h_*}{h} \right)^n - \left( \frac{h_*}{h} \right)^m \right], \qquad
     {\rm K} = \frac{A_{\rm H} L}{6\pi \epsilon \gamma_{\rm f} h_*^3 H^3}, \label{disjoining_pressure}
 \end{align}
with equilibrium thickness $\prm{h}_*$, exponents $n>m>1$  (we use $(n,m) = (3,2)$ since these values were shown by~\citet{lang13} to be appropriate for liquid metals), and Hamaker constant $A_{\rm H}$.
  We assume that the radiative heat loss from the film to the air is small compared to the heat conduction from the film to the substrate. As a consequence, we neglect heat loss through the liquid-air interface and apply Eq.~\eqref{system10}, an insulating boundary condition. Furthermore, we model the (assumed) primary heat loss mechanism through the interface between the film and the substrate at $z=0$ by perfect thermal contact via Eq.~\eqref{system11}. At $z=0$ we also assume no-slip and no-penetration of fluid via Eq.~\eqref{system12}. Finally, we assume that the film and substrate are thermally insulated at the lateral ends, $x=\pm N\pi$ and $y=\pm N\pi$.

 We will now proceed to simplify the full model as outlined above. We begin in \S\ref{Thermal section} with a discussion of the various models for heat conduction, and derive a leading order asymptotic model that (we will show) compares well with the full heat conduction model. In \S\ref{long_wave_section}, we discuss the long-wave approximation for thin films and the inclusion of thermal effects in the resultant thin film equation.
 
\subsection{Thermal Modeling}\label{Thermal section}

In what follows, we present three different models for the inclusion of thermal effects (the fluid dynamics in all cases will be described by the long-wave model, see \S\ref{long_wave_section}). In \S\ref{full_model_section} we give a ``Full" model for heat conduction, denoted (F), which includes both in-plane and out-of-plane heat diffusion, but omits both viscous dissipation and thermal advection. As discussed in \S\ref{intro_section}, a number of previous works have utilized a much simpler model that neglects lateral heat diffusion (e.g. \citet{trice_prb07,Dong_prf16,Seric_pof2018}). Although the exact relevance of such lateral (in-plane) heat transfer has not yet been carefully analyzed, prior work by \citet{Seric_pof2018} suggests that it may be important. To study and quantify the possible significance, in \S\ref{1d_model_section} we describe such a ``one dimensional" model for heat conduction, denoted (1D).  Finally, in \S\ref{asymptotic_model_section} we apply long-wave theory to (F) to develop an ``Asymptotic" model for heat conduction, (A). This model utilizes key assumptions on the non-dimensional parameters introduced in Eq.~\eqref{nd_parameters} to arrive at a system that is simpler than (F), but unlike (1D) retains lateral heat diffusion. These three models will be compared in \S\ref{Model Comparison Section}.

\subsubsection{Full Model}\label{full_model_section}
In order to compare models of heat conduction, we must first declare a model that serves as a benchmark. We refer to  Eqs.~\eqref{system5}--\eqref{system6}, \eqref{system10}--\eqref{system11}, and \eqref{system13}--\eqref{system15} (despite the presence of terms that may appear asymptotically small with respect to $\epsilon$ in comparison to other terms), as the \textbf{Full Model (F)} for heat conduction.

\subsubsection{1D Model}\label{1d_model_section}

Here we display the model obtained by neglecting in-plane heat conduction in (F), assuming that the term $\epsilon^2 \nabla_2^2 T_{\rm f}$ may be neglected compared with $\partial_z^2 T_{\rm f}$ in Eq.~\eqref{system5} but retaining all other terms. Equation \eqref{system10} is replaced by $\partial_z T_{\rm f}=0$ since $\boldsymbol{n}=(0,0,1)+O(\epsilon^2)$. This yields the following \textbf{1D Model (1D)} for heat conduction:
\begin{align}
    \epsilon^2 \mbox{Pe}_{\rm f} \partial_t T_{\rm f} &= \partial_z^2 T_{\rm f} +  \epsilon^2 Q, \label{thermal3:1} \quad &&\mbox{for\ }\quad z \in \left(0,h \right), \\
     \mbox{Pe}_{\rm s} \partial_t T_{\rm s} &= \partial_z^2 T_{\rm s}, \label{thermal3:2} \quad &&\mbox{for\ }\quad z \in \left(-H_{\rm s},0 \right), \\
    \partial_z T_{\rm f} &= 0, \quad &&\mbox{on\ }\quad z=h, \label{thermal3:3} \\
    \partial_z T_{\rm f} &= \mathcal{K}\epsilon^2 \partial_z T_{\rm s}, \quad &&\mbox{on\ }\quad z=0, \label{thermal3:4} \\
    T_{\rm f} &= T_{\rm s}, \quad &&\mbox{on\ }\quad z=0 , \label{thermal3:5} \\
    \partial_z T_{\rm s} &= \mbox{Bi} \left( T_{\rm s} - \prm{T}_{\rm a} \right), \quad &&\mbox{on\ }\quad z=-H_{\rm s},  \label{thermal3:6} \\
   \partial_x T_{\rm f} &=0, \quad &&\mbox{on\ }\quad x=\pm N\pi, \label{thermal3:7} \\
   \partial_y T_{\rm f} &=0, \quad &&\mbox{on\ }\quad y=\pm N\pi, \label{thermal3:8}
\end{align}
where $Q$ is given by Eq.~\eqref{Q_eqn}. We note that although the substrate temperature $T_{\rm s}$ only diffuses in the out-of-plane direction, $z$, it is still functionally dependent on the in-plane coordinates $x,y$, due to Eq.~\eqref{thermal3:5} and the dependence of film temperature $T_{\rm f}$ on $x,y$ (via dependence on film height $h$). It follows from \eqref{thermal3:5}, \eqref{thermal3:7} and \eqref{thermal3:8} that $\partial_x T_{\rm s}=0$ at $x=\pm N\pi$, and $\partial_y T_{\rm s}=0$ at $y=\pm N\pi$ automatically. For the rest of the paper, we refer to Eqs.~\eqref{thermal3:1}--\eqref{thermal3:8} as the (1D) model.

\subsubsection{Asymptotic Model}\label{asymptotic_model_section}

Next, we formulate a model of intermediate complexity by carrying out further asymptotic analysis. To do so, we first make a number of assumptions about the non-dimensional parameters defined in Eq.~\eqref{nd_parameters} and provide estimates of time scales based on the parameters given in Table \ref{table:ref_paras}:

\begin{enumerate}[(i)]
    \item  ${\rm Pe}_{\rm f}=O(1)$. The term $\epsilon^2 {\rm Pe}_{\rm f} = \left[ \left(\prm{\rho} \prm{c}\right)_{\rm f} \prm{H}^2/\prm{k}_{\rm f} \right]/\prm{t}_{\rm s}= t_{{\rm D}_{\rm f}}/\prm{t}_{\rm s}$ appearing in Eq.~\eqref{system5} is a ratio of two time scales: $t_{{\rm D}_{\rm f}}$, the time scale of diffusion of heat in the film, and $t_{\rm s}$, the time scale of film evolution. Thus, we assume $t_{{\rm D}_{\rm f}}\ll t_{\rm s}$; heat diffuses rapidly through the film, before any significant film evolution can occur. In our setup, $  t_{{\rm D}_{\rm f}}\approx 1.17~{\rm ps}$, whereas $\prm{t}_{\rm s} \approx 26.86~{\rm ns}$.
    \item ${\rm Pe}_{\rm s}=O(1)$. Similar to (i) the Peclet number for the solid layer can be written as a ratio of time scales, ${\rm Pe}_{\rm s}=\left[ \left(\prm{\rho} \prm{c}\right)_{\rm s} \prm{H}^2/\prm{k}_{\rm s} \right]/\prm{t}_{\rm s}= t_{{\rm D}_{\rm s}}/\prm{t}_{\rm s}$, where $t_{D_{\rm s}}$ is the time scale of out-of-plane thermal diffusion in the substrate. 
        We assume that this diffusion occurs on a time scale comparable to that of film evolution. Here, $t_{{\rm D}_{\rm s}}\approx 0.147$~ns. Although this is small relative to $t_{\rm s}$, this assumption ensures that the time-derivative is retained in Eq.~\eqref{thermal3:2}, which is numerically convenient and has a negligible effect on results.
    \item ${\rm Bi}=O(1)$. The Biot number ${\rm Bi}=(H/k_{\rm s})/(1/\alpha)$ can be interpreted as the ratio of internal thermal resistance due to diffusion, $H/k_{\rm s}$, and external thermal resistance, $1/\alpha$, due to convection away from the boundary $z=-H_{\rm s}$. We assume these internal and external thermal resistances are comparable.
    \item $\mathcal{K} =k_{\rm s}/(\epsilon^2 k_{\rm f}) = O(1)$; the film has much higher thermal conductivity than the substrate.
    \item $H_{\rm s} = O(1)$, indicating that the substrate thickness is comparable in size to the film thickness. Hence the substrate is also thin.
\end{enumerate}
The difference in length scales in the problem motivates the idea that in-plane and out-of-plane diffusion can occur on different time scales. As a consequence of the thin substrate assumption, (v), the in-plane diffusion is 
much slower than that of out-of-plane diffusion. The ratio of the film evolution time scale to that of diffusion is therefore much smaller for in-plane diffusion than out-of-plane. Consequently, in-plane diffusion can be neglected in the substrate (cf. \citet{Seric_pof2018}).

To obtain an asymptotically valid model we assume the following expansions:
\begin{align}
    T_{\rm f} = T_{\rm f}^{(0)} + \epsilon^2 T_{\rm f}^{(1)} + \cdots, \qquad
    T_{\rm s} = T_{\rm s}^{(0)} + \epsilon^2 T_{\rm s}^{(1)} + \cdots, \nonumber
\end{align}
so that,  on substituting in Eqs.~\eqref{system5}--\eqref{system6}, \eqref{system10}--\eqref{system11}, \eqref{system13}--\eqref{system14} and using assumptions (i)--(v) listed above, the leading order model is given by 
\begin{align}
    \partial_z^2 T_{\rm f}^{(0)} &=0,  \quad &&\mbox{for\ }\quad z\in(0,h), \label{lo_metal_temp} \\
  \mbox{Pe}_{\rm s} \partial_t T_s^{(0)} &= \partial_z^2 T_{\rm s}^{(0)} , \quad &&\mbox{for\ }\quad z\in(-H_{\rm s},0), \nonumber \\
  \partial_z T_{\rm f}^{(0)} &=0, \qquad &&\mbox{on\ }\quad z=h, \label{insulation_free_bound} \\
  \partial_z T_{\rm f}^{(0)} &=0, \qquad &&\mbox{on\ }\quad z=0, \label{cont_flux} \\
 T_{\rm f}^{(0)} &= T_{\rm s}^{(0)}, \qquad &&\mbox{on\ }\quad z=0, \nonumber \\
 \partial_z T_{\rm s}^{(0)} &= {\rm Bi} \left( T_{\rm s}^{(0)} - T_{\rm a} \right), \qquad &&\mbox{on\ }\quad z=-H_{\rm s}, \nonumber \\
 \partial_x T_{\rm f}^{(0)} &= 0, \qquad &&\mbox{on\ }\quad x=\pm N\pi, \nonumber \\
 \partial_y T_{\rm f}^{(0)} &= 0, \qquad &&\mbox{on\ }\quad y=\pm N\pi. \nonumber
\end{align}
Equations \eqref{lo_metal_temp}--\eqref{cont_flux} result in a leading order film temperature that is independent of $z$ but still unknown, $T_{\rm f}^{(0)}=T_{\rm f}^{(0)}(x,y,t)$. We must therefore proceed to next order in the asymptotic expansion to obtain a closed model for the leading order film temperature. 
Collecting terms at next order in Eq.~\eqref{system5} yields:
\begin{align}
    \mbox{Pe}_{\rm f} \partial_t T_{\rm f}^{(0)} = \nabla_2^2 T_{\rm f}^{(0)} + \partial_z^2 T_{\rm f}^{(1)} + F(t) \left[1-R(h) \right] \exp \left[- \alpha_{\rm f} \left( h-z \right) \right], \label{second_order_heat}
\end{align}
while the boundary conditions \eqref{system10} and \eqref{system11} at the same order are:
\begin{align}
    \partial_z T_{\rm f}^{(1)} &= \bnabla_2 h \cdot \bnabla_2 T_{\rm f}^{(0)}, \quad &&\mbox{on\ }\quad z=h, \label{next_order_flux_cont} \\
 \partial_z T_{\rm f}^{(1)} &= \mathcal{K} \partial_z T_{\rm s}^{(0)}, \quad &&\mbox{on\ }\quad z=0 \label{next_order_fs_cond}.
\end{align}
Since $T_{\rm f}^{(0)}$ is independent of $z$ we can integrate Eq.~\eqref{second_order_heat} from $z=0$ to $z=h$. Doing so, and applying the boundary conditions \eqref{next_order_flux_cont} and \eqref{next_order_fs_cond}, gives the following evolution equation for leading order film temperature:
\begin{align}
    h \mbox{Pe}_{\rm f} \partial_t T_{\rm f} = \bnabla_2 \bcdot \left( h \bnabla_2 T_{\rm f} \right) - \mathcal{K} \left(\partial_z T_{\rm s}\right)\vert_{z=0} + h \overline{Q}, \label{asymptotic model1}
\end{align}
for $x,y \in (-N\pi,N\pi)$, where $\overline{Q}=h^{-1} \int_{0}^{h} F(t) \left[1-R(h) \right] \exp \left[ -\alpha_{\rm f} \left(h-z \right) \right]dz$ is the averaged heat source and the superscripts on $T_{\rm f}$, $T_{\rm s}$ are dropped for convenience, since now only leading order quantities are considered. Here, $\bnabla_2 \bcdot \left( h \bnabla_2 T_{\rm f} \right)$ in Eq.~\eqref{asymptotic model1} describes the lateral heat diffusion, while the terms $\mathcal{K} \partial_z T_{\rm s}$ and $h \overline{Q}$ represent the heat lost from the film due to contact with the substrate and the generation of heat in the film due to the laser source, respectively. The final asymptotic model for heat conduction is Eq.~\eqref{asymptotic model1} in the film, together with:
\begin{align}
        \mbox{Pe}_{\rm s}\partial_t T_{\rm s} &= \partial_z^2 T_{\rm s}, \qquad &&\mbox{for\ }\quad z \in \left(-H_{\rm s},0 \right), \label{asymptotic_sub_eqn} \\
        T_{\rm f} &= T_{\rm s}, \qquad &&\mbox{on\ }\quad z=0, \label{cont_temp_nd} \\
        \partial_z T_{\rm s} &= {\rm Bi} \left( T_{\rm s} - T_{\rm a} \right), \qquad &&\mbox{on\ }\quad z=-H_s, \label{Bi_boundary_condition} \\
        \partial_x T_{\rm f} &=0, \qquad &&\mbox{on\ }\quad x=\pm N\pi, \\
        \partial_y T_{\rm f} &=0, \qquad &&\mbox{on\ }\quad y=\pm N\pi. \label{asymptotic_model_end}
\end{align}
 We note that, even though lateral diffusion is neglected in equation (\ref{asymptotic_sub_eqn}), the substrate temperature $T_{\rm s}$ remains a function of $x$, $y$ and $z$, the in-plane variation entering through the boundary condition \eqref{cont_temp_nd}. By the same reasoning as in the previous section, the lateral end insulating conditions on $T_{\rm s}$, Eqs.~\eqref{system14}--\eqref{system15}, are satisfied vacuously.

In summary, we have formulated an asymptotic model for heat conduction that exploits the natural geometry of the problem as well as the relative sizes of material parameters (assumptions (i)--(v)). This model, denoted (A), has advantages over both (F) and (1D). By integrating over the $z$-direction, a closed model is obtained for a leading order temperature profile that is independent of $z$, simplifying the problem significantly. As a consequence, (A) is considerably less computationally demanding than (F). Solving (F) for the temperature profile throughout the evolving film is cumbersome since the domain is deformable (see the appendix for details): model (A) eliminates this complication since film temperature depends only on the in-plane direction(s) and time. Model (A) is also (as we will see) substantially more accurate and faster to compute than (1D).

A number of other authors have developed reduced models for heat transfer within films, which we now briefly highlight and contrast with our model (A). The models presented by \citet{Dong_prf16,Seric_pof2018,trice_prb07} ignore in-plane diffusion in the substrate, similar to Eq.~\eqref{asymptotic_sub_eqn}. Furthermore, all use a Dirichlet boundary condition at the bottom of the substrate rather than the Newton law of cooling used here (Eq.~\eqref{Bi_boundary_condition}).  \citet{shklyaev12} arrive at a leading order temperature equation through arguments similar to ours above. Their model also retains the in-plane diffusion term, $\bnabla_2 \bcdot (h\bnabla_2 T_{\rm f})$ in Eq.~\eqref{asymptotic model1}, but considers radiative heat losses through the liquid-air interface to be dominant rather than the heat loss through the substrate. One important difference between our model (A) and that of \citet{shklyaev12} is that in (A) volumetric heating is considered, which depends on the local value of the film thickness. This fully couples the fluid and thermal problems, whereas the heating mode considered by \citet{shklyaev12} (heating from the substrate below) does not depend directly on the film thickness. 
\citet{atena09} also assume such volumetric heating but consider the case where the internal heat generation is promoted to leading order so that  $z$-dependence is retained in the film temperature,
leading to a more computationally demanding formulation.


\subsection{Free Surface Evolution}\label{long_wave_section}

Each of our heat conduction models couples to the film evolution problem, which must be solved simultaneously. Here we briefly summarize the long-wave approximation that we utilize in all our simulations, which effectively reduces the NS equations to a 4th order PDE for film thickness, $h$.  
To retain maximum generality and reasonable tractability, we allow both viscosity and surface tension (which appears in boundary conditions \eqref{system8}--\eqref{system9_2}) to vary with temperature but treat material density, specific heat, and thermal conductivity as fixed at their respective values at melting temperature. We present forms for both surface tension and viscosity that utilize the average free-surface temperature, defined for our purposes by
\begin{align}
    \overline{T}=\frac{1}{\left( 2N\pi \right)^2}\int_{-N\pi}^{N\pi} \int_{-N\pi}^{N\pi}(T_{\rm f}\vert_{z=h})\:\mathrm{d} x \mathrm{d} y. \label{T_bar_defn}
\end{align}

We assume that surface tension depends linearly on temperature in the following sense:
\begin{align}
    \gamma = 1 +\frac{2{\rm Ma}}{3} (\overline{T} - 1) +  \epsilon^2 \frac{2\mbox{Ma}}{3}  \Delta T + O(\epsilon^4) 
    = \Gamma + \epsilon^2 \frac{2\mbox{Ma}}{3} \Delta T + O(\epsilon^4), \label{surf_tens_dep}
\end{align}
where ${\rm Ma}$ is the Marangoni number, given by ${\rm Ma} = (3 \gamma_{\rm T} T_{\rm melt})/(2\gamma_{\rm f})$, where $\gamma_T=(\gamma_{\rm f}/T_{\rm melt})\mathrm{d}\gamma/\mathrm{d}\overline{T} \vert_{\overline{T}=1}$ is the change in surface tension with temperature when the film (on average) is at melting temperature, $\overline{T}=1$ (the factors of $2/3$ are used for later convenience); and $\Delta T$ is given by
\begin{align}
    \Delta T=T_{\rm f}\vert_{z=h}-\overline{T}. \label{Delta_T_eq}
\end{align}
Since $\overline{T}$ depends only on time, Eq.~\eqref{surf_tens_dep} can be interpreted as defining a surface tension that varies in time (at leading order) due to variations in the average temperature, and in space (at higher order), due to spatial variations in temperature. This asymptotic form of $\gamma$ proposed in Eq.~\eqref{surf_tens_dep} provides a consistent balance in the normal and tangential stress balances presented below. The temperature dependence of the dimensionless viscosity, $\mathcal{M}=\var{\mu}/\mu_{\rm f}$, is modeled by an Arrhenius-type relationship, which we take as
\begin{align}
    \mathcal{M} (t) &= \exp\left( \frac{\prm{E}}{\prm{R}\prm{T}_{\rm melt}}\left( \frac{1}{\overline{T}}-1 \right) \right), \label{viscosity_eq}
\end{align}
where $\prm{R}=8.314{\rm J} {\rm K}^{-1} {\rm mol}^{-1}$ is the universal gas constant and $\prm{E}$ is the activation energy \citep{metals_ref_book_2004}.

To leading order in $\epsilon^2$ the normal and tangential stress balances (Eqs.~\eqref{system8}--\eqref{system9_2}) are:
\begin{align}
    p &= - 3\Gamma \nabla_2^2 h - 3\Pi(h), \quad  &&\mbox{on\ }\quad z=h, \label{nsb_nd} \\
    \mathcal{M} \partial_z (u,v) &= 2{\rm Ma} \bnabla_2 \left( \Delta T \right), \quad &&\mbox{on\ }\quad z=h. \label{tsb_nd}
\end{align}
 To obtain an evolution equation for film thickness, we express conservation of mass in the form
\begin{align}
\partial_t h + \bnabla_2 \bcdot \left( h \overline{\boldsymbol{u}} \right)=0,  \label{revised_kbc} 
\end{align}
 where $\overline{\boldsymbol{u}}$ is the film-averaged (in-plane) velocity, $\overline{\boldsymbol{u}}=h^{-1} \int_{0}^{h} (u,v)\:\mathrm{d} z$. To determine $u$ and $v$, we expand pressure and velocity fields in Eqs. \eqref{system1}--\eqref{system3} to leading order in $\epsilon$, assume $\Gamma$ is $O(1)$, and apply the boundary conditions \eqref{nsb_nd} and \eqref{tsb_nd}, together with the kinematic condition \eqref{system7}, to obtain the leading order velocity profile,
 \begin{align}
     \left(u,v \right) = \frac{1}{\mathcal{M}}\left[ \left( \frac{z^2}{2}-zh \right)\bnabla_2 p + 2z {\rm Ma}\bnabla_2 \left( \Delta T \right) \right], \label{u_eqn}
 \end{align}
and $z$-independent pressure, $p$. Equation  \eqref{nsb_nd}, therefore, gives the pressure throughout the layer and $\bnabla_2 p$ is found by taking the gradient of \eqref{nsb_nd}. After plugging \eqref{u_eqn} into \eqref{revised_kbc} we then arrive at the thin film equation,
\begin{align}
    \partial_t h + \bnabla_2 \cdot \left[\frac{1}{\mathcal{M}} \Big(  h^3  \bnabla_2 \left( \Gamma \nabla_2^2 h + \Pi(h) \right) + h^2 {\rm Ma} \bnabla_2 \left( \Delta T \right)  \Big) \right]  = 0. \label{thin_film}  
\end{align}
Following the time derivative term in Eq.~\eqref{thin_film}, the terms (from left to right) represent the capillary, disjoining pressure, and Marangoni terms, respectively. In general, Eq.~\eqref{thin_film} describes the evolution of a nanoscale thin film and is applicable for any of our three thermal models (A), (F), or (1D) by using $\Delta T$ (Eq.~\eqref{Delta_T_eq}) and $\overline{T}$ (Eq.~\eqref{T_bar_defn}) from the appropriate model. 

Equation \eqref{thin_film} is already sufficiently general to incorporate in-plane variation of viscosity. 
For model (A), this may be accomplished by using $T_{\rm f}^{(0)}$ in place of $\overline{T}$ in equation \eqref{viscosity_eq}. This is an additional advantage of (A) that is not immediately shared by (F) or (1D) (including spatial dependence of viscosity is more complex with these models due to the dependence of temperature on $z$, as discussed further in the next section).

Although the choice of scales made at the start of \S\ref{model_formulation_section} is standard in the long-wave approximation (e.g. \citet{oron_rmp97}), the introduction of heat conduction adds significant complications, and it is important to check for consistency. For example, to retain surface tension to leading order in Eq.~\eqref{thin_film}, the velocity scale must be such that $\Gamma=O(1)$. This leads to the specific choice of time scale $t_{\rm s}$, which may be slower than the (nanoseconds) duration of the Gaussian pulse. 
Further discussion of the choice of scales is provided in \S\ref{appendix_scales}. We note that as a consequence of our chosen scalings, it would be asymptotically consistent to replace Eq.~\eqref{asymptotic model1} and Eq.~\eqref{asymptotic_sub_eqn} by their quasi-steady analogues; however, solving the resulting 
boundary value problems is numerically more complicated (and does not affect results), hence we retain these time derivatives in the formulation. 


\section{Results} \label{results_section}
 
 For simplicity, we limit our considerations to two spatial dimensions, eliminating $y$-dependence, so that the film's free surface is at $z=h(x,t)$. In \S\ref{lsa_section}, we perform linear stability analysis (LSA), which provides a framework for describing instability growth and motivates our choice of initial film profile(s). \S\ref{simulation_setup_section} outlines the setup of the simulations, including the initial conditions and numerical procedures. \S\ref{Model Comparison Section} and \S\ref{Experimental_comparison_section} show simulation results for both film and thermal evolution. In \S\ref{Model Comparison Section} we compare the thermal models. 
 In \S\ref{Experimental_comparison_section} we (almost) exclusively use (A) to solve for heat conduction and allow the surface tension and viscosity to vary with temperature. For what follows, we define the spatially-averaged film temperature,
 \begin{align}
  T_{\rm avg}=\frac{1}{2N\pi}\int_{-N\pi}^{N\pi} \frac{1}{h}\int_{0}^{h}T_{\rm f}\:\mathrm{d} z \mathrm{d} x, \label{T_avg_defn}
 \end{align}
where the film temperature $T_{\rm f}$ is found using model (1D), (F) or  (A) (leading-order temperature
for (A)).\footnote{For model (A) this is exactly the 2D free-surface average given by \eqref{T_bar_defn}. For models (F) and (1D) it is the average temperature of the entire film.} The parameters used are as given in Table \ref{table:ref_paras}, except where specified otherwise.
\subsection{Linear Stability Analysis (LSA)}\label{lsa_section}

To provide insight into the mechanism by which films dewet, we carry out linear stability analysis (LSA). Consider a uniform film of height $h_0$, perturbed as follows
 \begin{align}
     h(x,t)=h_0 \left( 1 + \delta e^{ikx + \beta t} \right), \nonumber
 \end{align}
where $k$ is the wavenumber, $\beta$ is the growth rate, and $\delta \ll 1$ is the amplitude. A more complete analysis could also incorporate independent perturbations to temperature profiles, as was done by \citet{shklyaev12}; for simplicity we do not take this approach. We also neglect, for now, the influence of thermal gradients on film instability by setting ${\rm Ma}=0$ in Eq.~\eqref{thin_film}. LSA then provides the following dispersion relation, 
\begin{align}
    \beta (k) &= \frac{1}{\mathcal{M}}h_0^3 k^2 \left( 2 -  \Gamma k^2 \right). \label{dispersion:1}
\end{align}
From Eq.~\eqref{dispersion:1}, it is immediately apparent that viscosity sets the timescale of the perturbation growth/decay. The stability of the film, on the other hand, is controlled by the surface tension. For our purposes we only consider perturbations that grow ($\beta>0$ when $k^2<2/\Gamma$). The wavenumber $k_{\rm m}$ corresponding to maximum growth is found from Eq.~\eqref{dispersion:1} by setting $\partial \beta/\partial k=0$. The wavelength of maximum growth $\Lambda_{\rm m}$ and the maximum growth rate $\beta_{\rm m}=\beta(k_{\rm m})$ can then be written in the simple form:
\begin{align}
    \Lambda_{\rm m} = 2\pi \sqrt{\Gamma}, \qquad \beta_{\rm m}\equiv \frac{h_0^3}{\Gamma \mathcal{M}}. \label{dispersion_lambda_and_beta}
\end{align}
  Since increasing temperature decreases $\Gamma$ and $\mathcal{M}$ (see Eqs.~\eqref{surf_tens_dep} and \eqref{viscosity_eq}), higher temperatures will lead to smaller $\Lambda_{\rm m}$ and larger $\beta_{\rm m}$. In what follows next, $\Lambda_{\rm m}$ will be used to define simulation geometries. We note that  $\Lambda_{\rm m}$ and $L$ are related when $\Gamma=1$ via $\Lambda_{\rm m}=\lambda_{\rm m}/L = 2\pi$, where the expression for $\lambda_{\rm m}$ is given in \S\ref{appendix_lsa}.

\subsection{Simulation Setup}\label{simulation_setup_section}
Here we describe the details of the simulations.  The numerical solution of \eqref{thin_film} is obtained using an approach adapted from \citet{DK_jcp02} with uniform grid size, $\Delta x=h_*$ ($h_*$ is defined in Eq.~\eqref{disjoining_pressure} and is fixed for all simulations as $h_*=0.1$), which is sufficient to ensure accuracy. Models (F), (1D), and (A) are all solved using central difference spatial discretisation. Model (F) utilizes a mapping onto a rectangle to account for the moving boundary (this is not needed for models (1D) and (A) due to lack of in-plane diffusion and lack of $z$-dependence, respectively); see \S\ref{numerical_F_section} in the Appendix for details. For (1D) temporal discretisation is performed using the Crank-Nicolson scheme, while for (A) an implicit-explicit (IMEX) scheme is used (see \S\ref{numerical_A_section}). Model (F) is solved using an alternating direction implicit (ADI) method, treating mixed derivative terms explicitly. Adaptive time stepping is used to ensure a tolerance of $10^{-3}$ maximum allowable relative error in temperature and film thickness. Note that the time-stepping tolerance criteria must be satisfied for both film and heat evolution equations in order to proceed with a successful iteration (a point to which we return later). 
No-flux boundary conditions $\partial_x h=\partial_x^3 h=0$ are imposed at $x=\pm N\pi$ ($h\overline{\boldsymbol{u}}=0$ from Eq. \eqref{revised_kbc}). The domain length, $2N\pi = N\Lambda_{\rm m}(\Gamma=1)$, is now set by fixing $N=1$ or $N=20$. For $N=1$, the initial film profile is set to represent a small perturbation to a uniform film thickness,
\begin{align}
    h(x,0) &= h_0 \left( 1 - \delta \cos \left( \frac{2 \pi x}{\Lambda_{\rm m}} \right) \right), \label{h_ic1}
\end{align}
where $\delta=0.01$. We refer to the corresponding simulations as those with domain length $\Lambda_{\rm m}$. For $N=20$ the following initial film profile is imposed:
\begin{align}
    h(x,0) = h_0 \left[ 1 + \delta \sum\limits_{i=1}^{40} A_i \Big( \cos \left( 2\pi x/\lambda_i \right) + \sin \left( 2\pi x/\lambda_i \right) \Big) \right], \label{h_random_pert}
\end{align}
where the amplitudes $A_i$ are randomly chosen in $[-1,1]$ and $\lambda_i = 2\Lambda_{\rm m}/i$. Similarly, we refer to simulations that use Eq.~\eqref{h_random_pert} as those with domain length $20\Lambda_{\rm m}$. For 
both values of $N$, the film and substrate are each initially set to the ambient temperature,
\begin{align}
    T_{\rm f}^{(0)}(x,0)=T_{\rm f}(x,z,0)= T_{\rm s}(x,z,0) = T_{\rm a}. \nonumber
\end{align}

The numerical solutions for $T_{\rm f}$ and $T_{\rm s}$ are found first, with the film static, since the film is initially solid ($T_{\rm a}<1$). Once the film is melted (we define this shortly) the solutions for $h$, $T_{\rm f}$, $T_{\rm s}$ are then iterated successively. The flow of the numerical algorithm is as follows:
\begin{itemize}
    \item \textbf{Film solid and static}
    \begin{itemize}
        \item Update $T_{\rm f}$
        \item Update $T_{\rm s}$
        \item Repeat the previous 2 steps until melted.
    \end{itemize}
    \item \textbf{Film melted}.
    \begin{itemize}
        \item Set $\Gamma, \mathcal{M}, \Delta T$. Update $h(x,t)$
        \item Update $T_{\rm f}$
        \item Update $T_{\rm s}$
        \item Repeat the previous 3 steps until re-solidification.
    \end{itemize}
    \item \textbf{End}.
\end{itemize}
Once the film is melted, both film evolution and heat conduction are solved at every time step (although the numerical algorithm allows for a less frequent numerical solution of the temperature equation relative to that of $h$). The film evolution is coupled to the temperature profile through the material parameters $\Gamma$ (surface tension) and $\mathcal{M}$ (viscosity), and the Marangoni term $\Delta T$ in Eq.~\eqref{thin_film}. The film is allowed to evolve (flow) only when the temperature is everywhere greater than the melting temperature: it then evolves according to Eq.~\eqref{thin_film}. Subsequently, as the laser heat source decays, the film temperature eventually drops below the melting point and the film re-solidifies. All simulations shown in this paper are ended when the average temperature decreases to solidification temperature, $T_{\rm avg}=1$. In what follows, we will be using the liquid lifetime (LL), defined as the time interval during which the average film temperature is above melting ($T_{\rm avg}>1$).

\subsection{Model Comparison with Fixed Parameters} \label{Model Comparison Section}

We now compare  models (F), (1D), and (A) holding the material parameters fixed. As a basic check, we first consider a stationary flat film  ($h=h_0$), with material parameters fixed at the values corresponding to the melting temperature ($\Gamma=1 ,\mathcal{M}=1$). For
such a film there is no in-plane heat conduction: the temperature $T_{\rm f}$ is a function of time $t$ only; and models (F), (1D), and (A) all agree. 
\begin{figure}
     \centering
    \includegraphics[width=\textwidth]{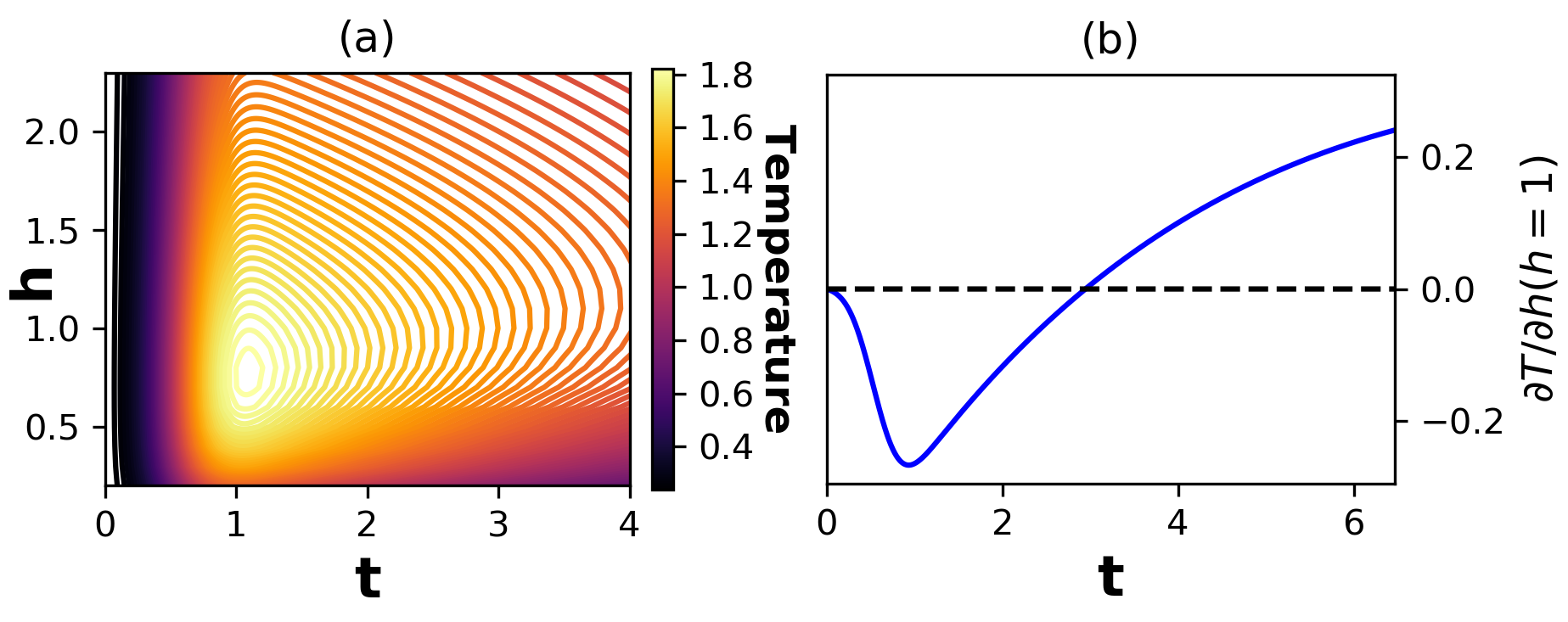}
    \caption{(a) Contour plots of average film temperature for a static flat film. (b) Rate of change of temperature with film thickness, $\partial T_{\rm f}/\partial h$ as a function of time for $h=1$. At early times ($t<2.95$) $\partial T_{\rm f}/\partial h<0$ and later ($t>2.95$) $\partial T_{\rm f}/\partial h>0$.}
    \label{fig:flat_film_temp_contours}
\end{figure}

Figure \ref{fig:flat_film_temp_contours}a plots average film temperature against $h_0$ and time, showing that temperature depends on film thickness in a non-monotonous manner. Figure \ref{fig:flat_film_temp_contours}b plots the change of temperature with film thickness, $\partial T_{\rm f}/\partial h$, evaluated at $h=1$ (this value of $h$ will be used in later simulations), as a function of time. 
For early times ($t<2.95$), $\partial T_{\rm f}/\partial h<0$, so that a decrease in film thickness corresponds to an increase in temperature (thinner film is hotter). For later times $\partial T_{\rm f}/\partial h>0$ so that a decrease in film thickness leads to a decrease in temperature (thinner film is colder). This non-monotonic behavior of $\partial T_{\rm f}/\partial h$ in time is due to the changing balance between the heating from the source and the heat loss through the substrate (see \S\ref{appendix_dTdh} for more details).

\begin{figure}
    \centering
    \includegraphics[width=\textwidth]{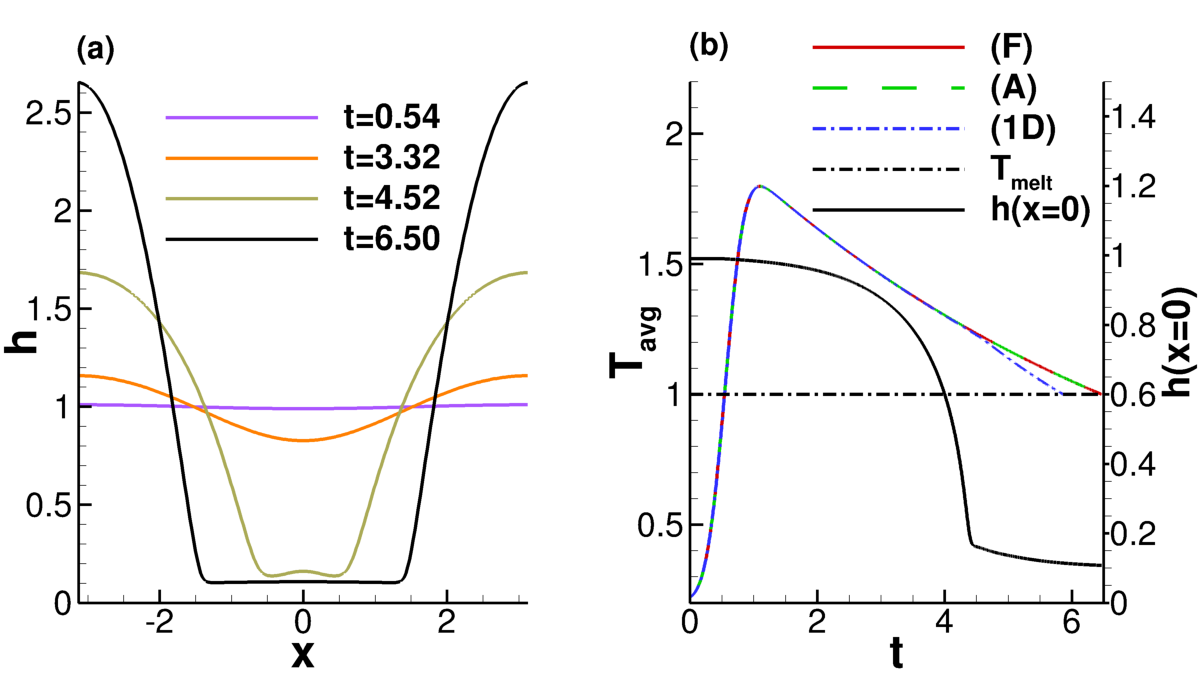}
    \caption{
        (a) Evolution of film thickness when material parameters are fixed and $h(x,0)$ is given by Eq.~\eqref{h_ic1}, at a few representative time points.
        (b) Average film temperature (see Eq. \eqref{T_avg_defn}) and midpoint film thickness $h(0,t)$ for the film profiles given in (a).
        Deviation between the models appears after the the film dewets. The material parameters are set to their melting temperature values, $\Gamma=1$ and  $\mathcal{M}=1$.}

    \label{fig:thickness_and_avg_temp_melt_visc}
\end{figure}

We next consider evolving films, with
the initial film profile given by Eq.~\eqref{h_ic1}.
Figure~\ref{fig:thickness_and_avg_temp_melt_visc}a shows the evolution using model (F), though the behaviour is also representative of (A) and (1D). Melting temperature is reached at $t\approx 0.54$; by time $t=3.32$ the liquid film begins to evolve appreciably; at $t=4.52$ significant 
 film evolution has occurred and the film first reaches the equilibrium film thickness; and at $t=6.50$ the film 
 has fully dewetted. Figure~\ref{fig:thickness_and_avg_temp_melt_visc}b shows average film temperatures using models (F), (1D) and (A), as well as the film height at the midpoint $x=0$. The average film temperatures are in good agreement before dewetting, but afterwards model (1D) begins to deviate significantly from models (F) and (A), which show excellent agreement for the entirety of the simulation. The cooling rate, $\mathrm{d}T_{\rm avg}/\mathrm{d}t$, is faster for (1D) than for (F) and (A) since heat cannot diffuse laterally through the film in (1D). This, in turn, produces a film that solidifies sooner (this will be discussed further in \S\ref{Experimental_comparison_section}). Despite this difference, the midpoint film height $h(x=0)$, shown here for (F), is similar for all models (see also Fig.~\ref{fig:thickness_and_avg_temp_melt_visc}a).


\begin{figure}
    \centering
    \includegraphics[width=\textwidth]{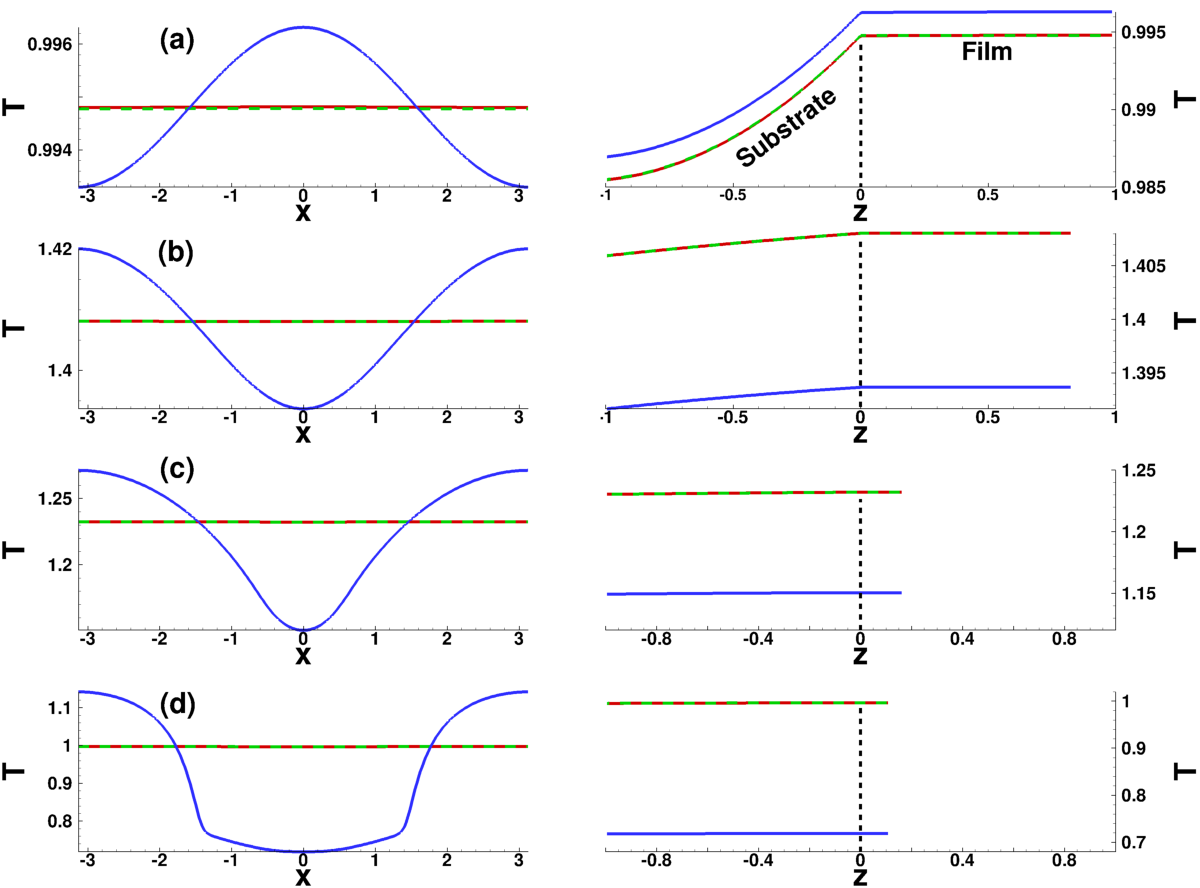}
    \caption{
    Left: Free surface ($z=h$) temperature profiles, $T_{\rm f}\vert_{z=h}$ and Right: Midpoint ($x=0$) temperature corresponding to times (a) $t=0.54$, (b) $t=3.32$, (c) $t=4.52$, and (d) $t=6.50$. Here $\Gamma=1$, $\mathcal{M}=1$ and the overlapping curves correspond to models (F) and (A).  Note the difference in vertical axis scales between parts (a)-(d). Color code: (F) ({\color{red} red}),  (A) ({\color{green} green} dashed), (1D) ({\color{blue} blue}). }
    \label{fig:Tfilm_and_Tsub_full}
\end{figure}

 We next discuss the spatial variation of temperature. Figure \ref{fig:Tfilm_and_Tsub_full} shows free surface temperatures (left column) and film midpoint temperatures (at $x=0$; right column) at the times displayed in Fig.~\ref{fig:thickness_and_avg_temp_melt_visc}a. Since the film thins at its midpoint as $t$ increases (see Fig. \ref{fig:thickness_and_avg_temp_melt_visc}b), the film domain ($z>0$) shrinks from (a)-(d) in the right column.
The lateral spatial variation of temperature is seen to be much weaker for models (F) and (A) than for model (1D) (left column), due to the inclusion of lateral heat diffusion in (F) and (A); and the substrate/film midpoint temperatures are lower for (1D) than for (F)/(A) in frames (b)-(d) (right column).
In particular we find that the temperature predictions of models (F) and (A) differ by at most $0.01$\%, whereas (F) and (1D) differ by as much as 30\%. For model (1D) the temperature is initially higher at the film midpoint ($x=0$) than at the edges, a situation that is reversed at later times (e.g. Fig.~\ref{fig:Tfilm_and_Tsub_full}b). Using model (1D), therefore, may lead to overestimated temperature gradients, such as in \citet{trice_prb07}, which may significantly alter the evolution of the film.  

\begin{figure}
    \centering
    \includegraphics[width=\textwidth]{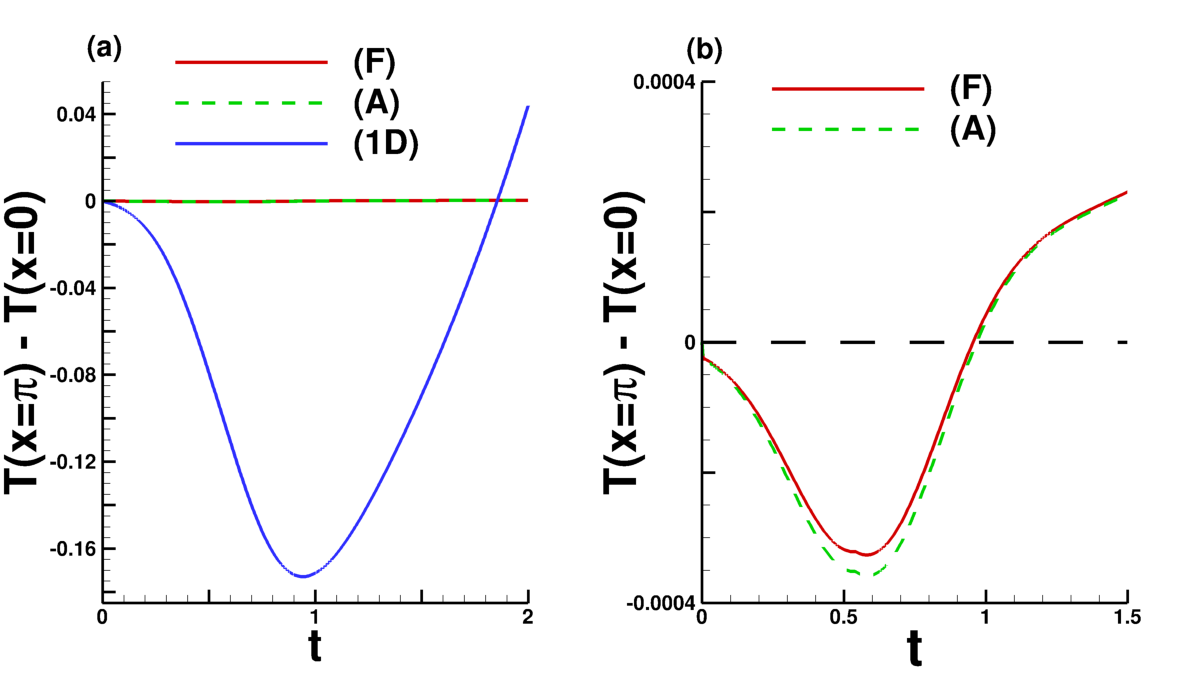}
    \caption{(a) Difference between free surface temperatures at the thickest part of the film $(x=\pm \pi$)
 and the thinnest part $(x=0)$ for models (F) ({\color{red} red}), (A) ({\color{green} green}), and (1D) ({\color{blue} blue}).
 (b) Zoom-in of results for models (F) and (A) from (a) to illustrate behaviour more clearly. The black dashed line represents the horizontal axis where $T(x=\pi)=T(x=0)$.}
    \label{fig:free_surf_hot_vs_cold}
\end{figure}

 To emphasize the lateral temperature variation  for all models, Fig.~\ref{fig:free_surf_hot_vs_cold} plots the difference between the free surface temperatures at the thinnest ($x=0$) and thickest ($x=\pm \pi$) parts of the film.
 Consistent with Fig.~\ref{fig:Tfilm_and_Tsub_full}a this temperature difference for model (1D) is much larger than for models (F) and (A).
 All models show the same trend: initially the thinnest part of the film is hottest, but ultimately the thickest parts are hottest. We attribute this change of behaviour,
 which occurs relatively early in the film evolution, to the combination of lateral diffusion and the heat loss through the substrate (see Eq. \eqref{cont_flux}).

 To conclude this section, we summarize our main findings. We have compared models (F), (1D), and (A) and found that (A) provides a much better approximation to (F) than does (1D). After dewetting, the film cools more rapidly with model (1D) than with (A) and (F) due to the neglect of in-plane thermal diffusion. Consequently, the average film temperatures in model (1D) vary significantly from those predicted by models (F) and (A).
 From a computational point of view, since the film temperature is independent of $z$ in model (A), it is significantly more computationally efficient than (F), and even more efficient than (1D). 
 For illustration, we note that for a $255 \times 200$ computational grid in $x$ and $z$, the simulation times for a typical run reported in this section are $73.4$, $1.3$, $4.8$ hours for models (F), (A), and (1D) respectively, on a 
 reasonably fast workstation. 


\subsection{Variation of Material Parameters} \label{Experimental_comparison_section}
In this section we consider the effect of varying surface tension and viscosity with temperature (see Eqs.~\eqref{surf_tens_dep} and \eqref{viscosity_eq}) and consider the influence of the Marangoni effect, by comparing film evolution with ${\rm Ma}=0$ and ${\rm Ma}\neq 0$. In the previous section (as well as in additional tests, not reported here for brevity),
we have demonstrated that model (A) provides a good approximation to the full model (F) with considerably less computational effort, and so henceforth, we will focus on exploring the differences between models (A) and (1D). 
In this section domain lengths of $\Lambda_{\rm m}$ and $20\Lambda_{\rm m}$ are simulated, beginning with the former (domain length of $\Lambda_{\rm m}$ may be assumed until otherwise stated).

\begin{figure}
    \centering
    \includegraphics[width=\textwidth]{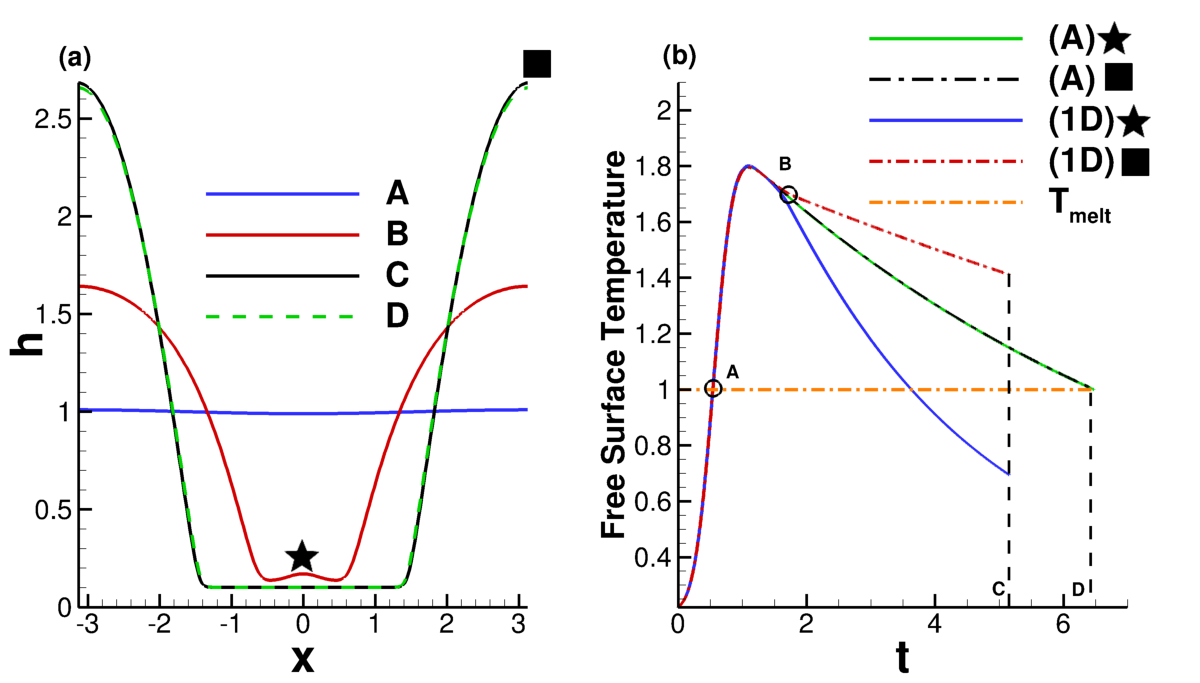}
    \caption{(a) Film thickness profiles (simulated with (F), but representative of (A) and (1D) also) for times \textbf{A} (melting), \textbf{B} (dewetting), \textbf{C} (1D) re-solidification and \textbf{D} (A) re-solidification. The markers in (a) represent $x=0$ ($\bigstar$) and $x=\pi$ ($\blacksquare$). (b) Free surface temperature at $\bigstar$ and $\blacksquare$ for (A) and (1D). The temperature profiles agree until the film dewets (\textbf{B}). Then, (1D) temperatures vary significantly at $\bigstar$ and $\blacksquare$, whereas (A) produces similar temperatures at both locations. Surface tension and viscosity vary in time, but Marangoni effect is not included (${\rm Ma}=0$).}
    \label{fig:thickness_and_temp_at_x0}
\end{figure}

We focus first on the case where both surface tension and viscosity depend on average temperature and are therefore time-dependent, $\Gamma(t),\mathcal{M}(t)$, but we ignore the Marangoni effect (set ${\rm Ma} = 0$ in Eq.~\eqref{thin_film}). In the subsequent text any reference to time-dependent surface tension or viscosity, $\Gamma(t),\mathcal{M}(t)$, refers solely to time-dependence through the average temperature. Figure \ref{fig:thickness_and_temp_at_x0}a shows the film thickness profiles and Fig.~\ref{fig:thickness_and_temp_at_x0}b shows the free-surface temperature profiles, $T_{\rm f}\vert_{z=h}$, at the thin ($\bigstar$) and thicker ($\blacksquare$) parts of the film. 
We observe that 
model predictions differ only after dewetting (\textbf{B}) and therefore the film thickness profiles in (1D) and (A) remain nearly identical. After dewetting, the two models exhibit marked differences in the temperature profiles. The large difference in cooling rates between (A) and (1D) is consistent with Fig.~\ref{fig:thickness_and_avg_temp_melt_visc}b. 
For (1D) this leads to the thin part of the film $\bigstar$ being significantly colder than the thicker parts $\blacksquare$ and the difference is exacerbated by the retention of heat in the thick part of the film.

\begin{figure}
    \centering
    \includegraphics[width=\textwidth,trim={0.15cm 0 0.15cm 0},clip]{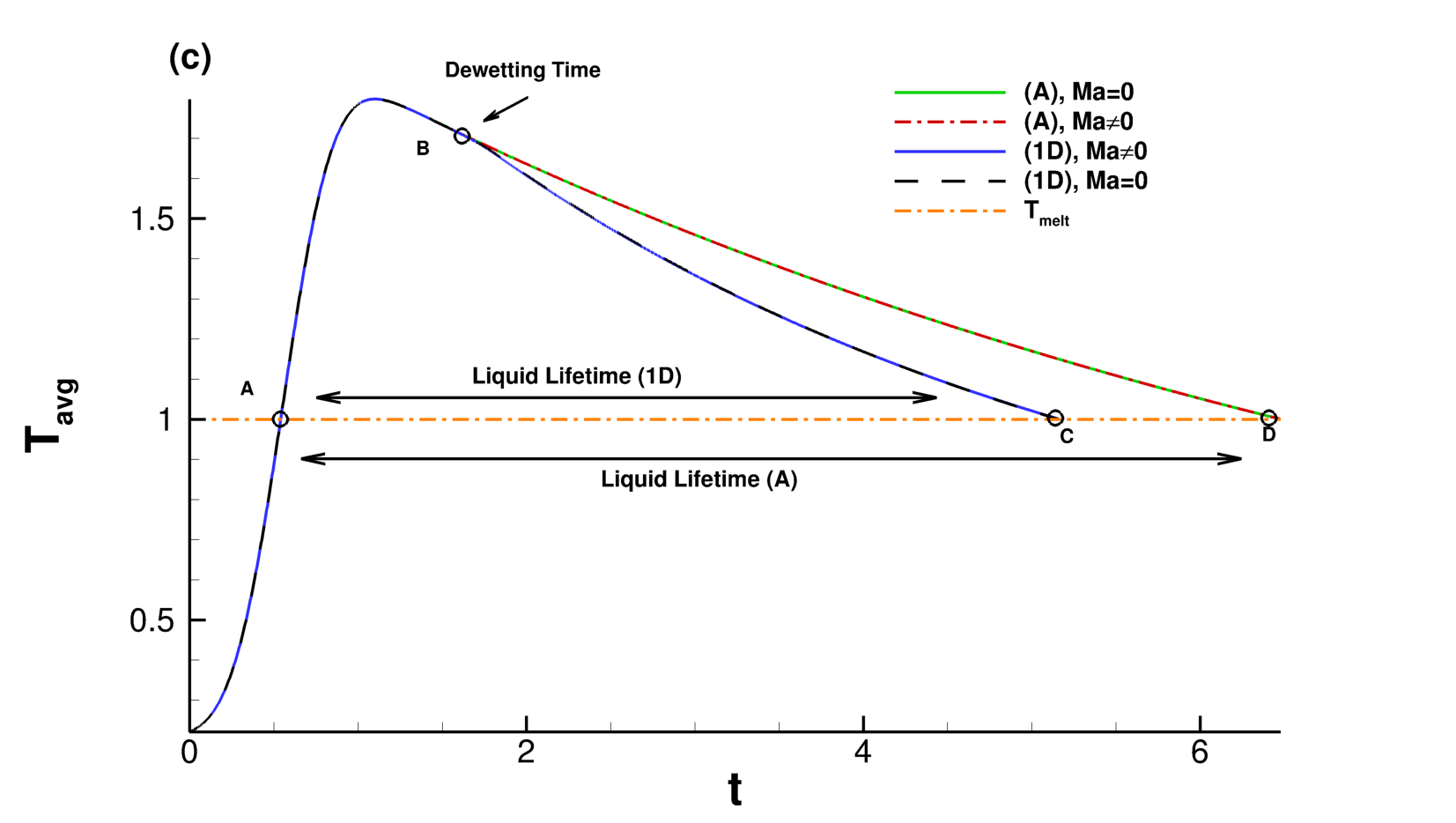}
    \caption{Average film temperatures using models (A) and (1D) for ${\rm Ma}=0$ and ${\rm Ma} \neq 0$,
    and using
 $\Gamma(t)$ and $\mathcal{M}(t)$. Points \textbf{A-D} correspond to those of Fig.~\ref{fig:thickness_and_temp_at_x0}(a). Near $t=0.54$ (\textbf{A}), the film melts in both models. The models begin to deviate around $t=1.72$ (\textbf{B}) when dewetting occurs; from this point until solidification (which for ${\rm Ma}=0$ occurs at \textbf{C} for (1D), \textbf{D} for (A); the times for the ${\rm Ma}\neq 0$ cases are similar) the temperature in model (1D) is lower than that in (A). The liquid lifetimes (LL) are ${\rm LL}\approx 4.73$ (1D) and ${\rm LL}\approx 5.94$ (A).
 }
    \label{fig:ll_comparison}
\end{figure}

Figure \ref{fig:ll_comparison} shows the average film temperatures for models (A) and (1D) with $\Gamma(t),\mathcal{M}(t)$ as considered in Fig.~\ref{fig:thickness_and_temp_at_x0}, for both ${\rm Ma}=0$ and ${\rm Ma}\neq 0$.
The Marangoni effect seems to have only a very minor influence on $T_{\rm avg}$ for both models.
Furthermore, the rapid cooling seen in model (1D) leads to a significantly shorter liquid lifetime (${\rm LL}=4.73$) than that predicted by model (A) (${\rm LL}=5.94$).

\begin{figure}
    \centering
    \includegraphics[width=\textwidth]{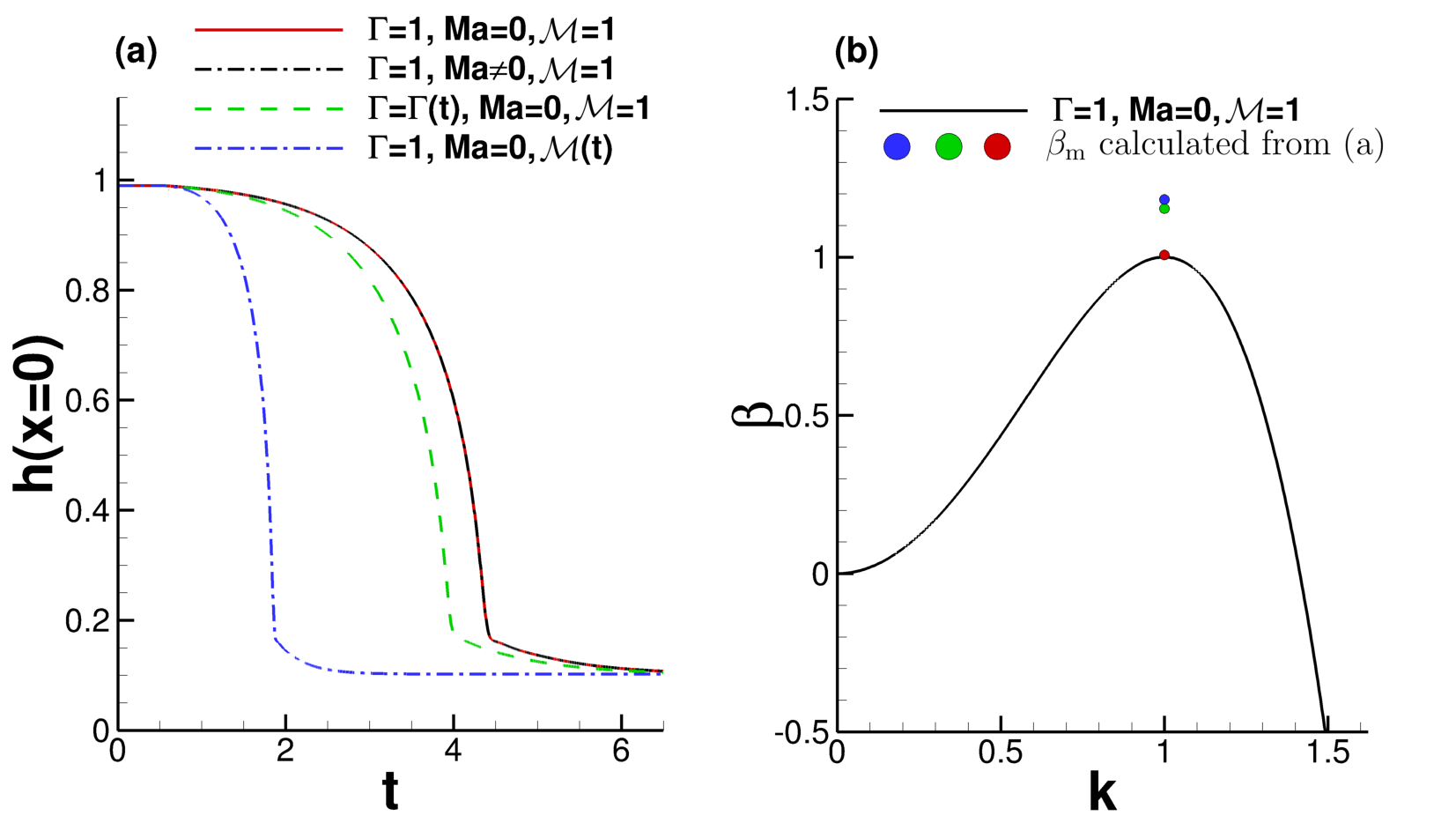}
    \caption{(a) Midpoint film thickness, $h(0,t)$, for the cases: surface tension and viscosity fixed, $\Gamma=1,\mathcal{M}=1$, with no Marangoni effect, $\mbox{Ma}=0$ ({\color{red} red}, solid line); surface tension and viscosity fixed with Marangoni effect, $\mbox{Ma}\neq 0$ (\textbf{black}, dash-dotted line); surface tension varies in time, $\Gamma(t)$, with viscosity fixed and no Marangoni ({\color{green} green}, dashed line); viscosity varies in time, $\mathcal{M}(t)$, surface tension fixed and no Marangoni ({\color{blue} blue}, dot-dashed line). In all cases model (A) was used to calculate temperature. (b) Growth rate, $\beta$ as a function of wavenumber, $k$, using Eq.~\eqref{dispersion:1} with $\Gamma=1$ and $\mathcal{M}=1$ and no Marangoni effect, ${\rm Ma}=0$. The {\color{blue} blue}, {\color{green} green}, and {\color{red} red} dots represent maximum growth rate extracted from the corresponding simulations in (a).}
    \label{fig:surf_and_marangoni_comparison}
\end{figure}
 
 Figure \ref{fig:surf_and_marangoni_comparison} compares the results obtained with(out) temperature
 variation of material properties.
From Fig.~\ref{fig:surf_and_marangoni_comparison}a we immediately conclude that
the Marangoni effect is very weak,
 and will be ignored henceforth (${\rm Ma=0}$ for the remaining results). The weak Marangoni effect is, at least in part, due to the high thermal conductivity of the film, which sets the thermal timescale and gives rise to the very weak spatial variations in interfacial temperatures seen in Fig.~\ref{fig:Tfilm_and_Tsub_full}.
 The second observation is that allowing surface tension to depend on time, $\Gamma=\Gamma(t)$, has a small but measurable effect on the results. With this dependence included, the film instability appears to develop faster than in the constant-$\Gamma$ case (compare the green dashed curve with the red solid curve in Fig.~\ref{fig:surf_and_marangoni_comparison}a).
 The third observation is that the time-dependence of viscosity has by far the largest effect on the film instability development, leading to much faster dewetting. 
 
 We  now  revisit  the  predictions  of  LSA and  compare them to simulation results. Figure \ref{fig:surf_and_marangoni_comparison}b plots the dispersion curve according to Eq. \eqref{dispersion:1}, for the constant parameter case $\Gamma=1$ and $\mathcal{M}=1$ (with $\mbox{Ma}=0$). To estimate the maximum growth rate, $\beta_{\rm m}$, in our numerical simulations, we assume the film perturbation grows exponentially at early times, consistent with LSA: $h=h_0(1+\delta\exp({ik_{\rm m}x+\beta_{\rm m} t}))$, where $k_{\rm m}$ is the corresponding wavenumber.\footnote{Though in practice perturbations of many different wavenumbers exist, the one usually most apparent in the unstable regime is $k_{\rm m}$.}  We then perform a best linear fit of $\ln((h(0,t)-h_0)/(\delta h_0))$ versus $t$ for early times. The {\color{red} red}, {\color{green} green} and {\color{blue} blue} dots correspond to growth rates $\beta_{\rm m}$ extracted from the corresponding colour-coded simulations in Fig.~\ref{fig:surf_and_marangoni_comparison}a. The film with parameters fixed ({\color{red} red}) grows at the rate predicted by LSA, whereas the film with time-dependent surface tension ({\color{green} green}) grows at a slightly faster rate. The growth rate in the time-dependent viscosity case ({\color{blue} blue}) is similar to the time-dependent surface tension case, despite the much faster instability development in Fig.~\ref{fig:surf_and_marangoni_comparison}a. This indicates the relevance of the nonlinear part of 
 instability growth.

\begin{figure}
    \centering
    \includegraphics[width=\textwidth,trim={0.2cm 0.1cm 0.1cm 0.1cm},clip]{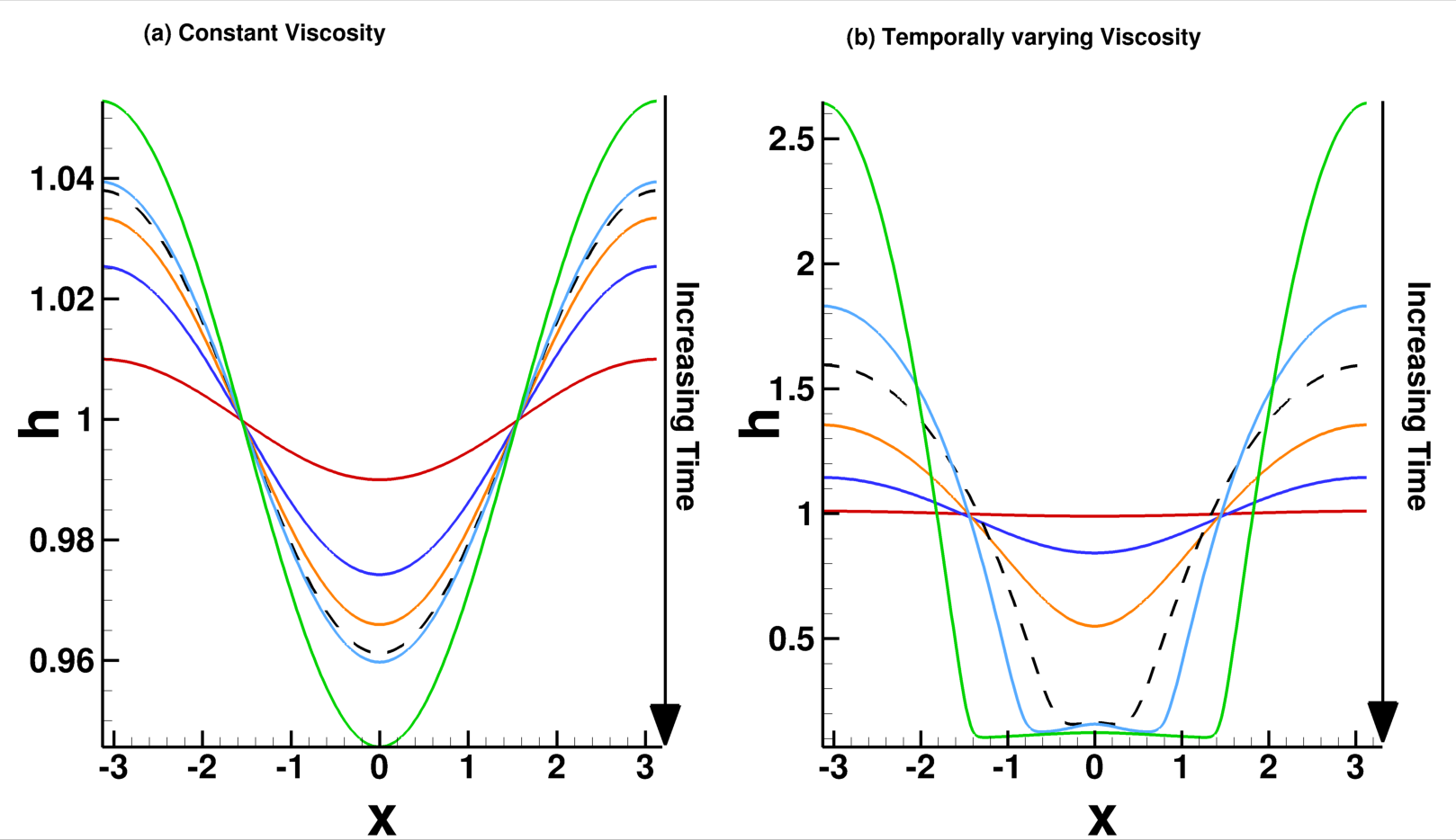}
    \caption{Comparison of free surface evolution $h(x,t)$ when (a) viscosity is fixed at the melting value, $\mathcal{M}=1$, and (b) viscosity varies in time according to average temperature, $\mathcal{M}(t)$ (see Eq. \eqref{viscosity_eq}). In both cases surface tension is fixed at the melting value $\Gamma=1$, and all times plotted are prior to film re-solidification. In this and in the figures that follow, ${\rm Ma}=0$. In (a) we find that re-solidification happens before dewetting (note the vertical axis scale), while the converse is true for (b), indicating the importance of variable viscosity. The times are as follows: $t=0$ ({\color{red} red}), $t=1.47$ ({\color{blue} blue}), $t=1.75$ ({\color{orange} orange}), $t=1.88$ (black dashed), $t=1.92$ ({\color{cyan} light blue}), $t=2.21$ ({\color{green} green}).}
    \label{fig:visc_comparison}
\end{figure}

 To highlight further the significance of time-dependent viscosity, Fig.~\ref{fig:visc_comparison} 
 compares film evolution for the constant (a) and variable (b) viscosity cases. The difference in film evolution is significant, with much larger instability growth rate for time-dependent viscosity.

To summarize, we have simulated and compared the results with and without the Marangoni effect, with and without time-varying surface tension, and with and without time-varying viscosity. We have found that allowing either surface tension or viscosity to depend on time through the average temperature speeds up the dewetting mechanism.  In particular, varying viscosity has the strongest impact on film dynamics.

For completeness we also consider the possible variation of viscosity with $x$. For model (A), since $T_{\rm f}^{(0)}(x,t)$ is a function only of the in-plane spatial variable $x$ (and time $t$), we may replace $\overline{T}$ in the definition of $\mathcal{M}$ (Eq.~\eqref{viscosity_eq}) by $T_{\rm f}^{(0)}$, so that viscosity, which we denote in this case by $\mathcal{M}(x,t)$, depends on both space and time, namely:
\begin{align}
        \mathcal{M} (x,t) &= \exp\left( \frac{\prm{E}}{\prm{R}\prm{T}_{\rm melt}}\left( \frac{1}{T_{\rm f}^{(0)}(x,t)}-1 \right) \right). \label{spatiotemp_viscosity_eq_A}
\end{align}
For model (1D) on the other hand, the calculated film temperature $T_{\rm f} (x,z,t)$ depends on both spatial coordinates $x$ and $z$, so obtaining an analogous model for $\mathcal{M}(x,t)$ requires some additional assumptions. We use the free-surface temperature $T_{\rm f}\vert_{z=h}$, so that viscosity takes the form
\begin{align}
        \mathcal{M} (x,t) &= \exp\left( \frac{\prm{E}}{\prm{R}\prm{T}_{\rm melt}}\left( \frac{1}{\left(T_{\rm f}\vert_{z=h}\right)}-1 \right) \right). \label{spatiotemp_viscosity_eq_1D}    
\end{align}

The film thickness profiles produced in these $\mathcal{M}(x,t)$ cases (both (1D) and (A)) are similar to the $\mathcal{M}(t)$ cases shown above and are thus omitted from the main text for brevity (an example is shown in Appendix \ref{appendix_high_act_energy}). 
\begin{figure}
    \centering
    \includegraphics[width=\textwidth,trim={0.1cm 0 0.1cm 0},clip]{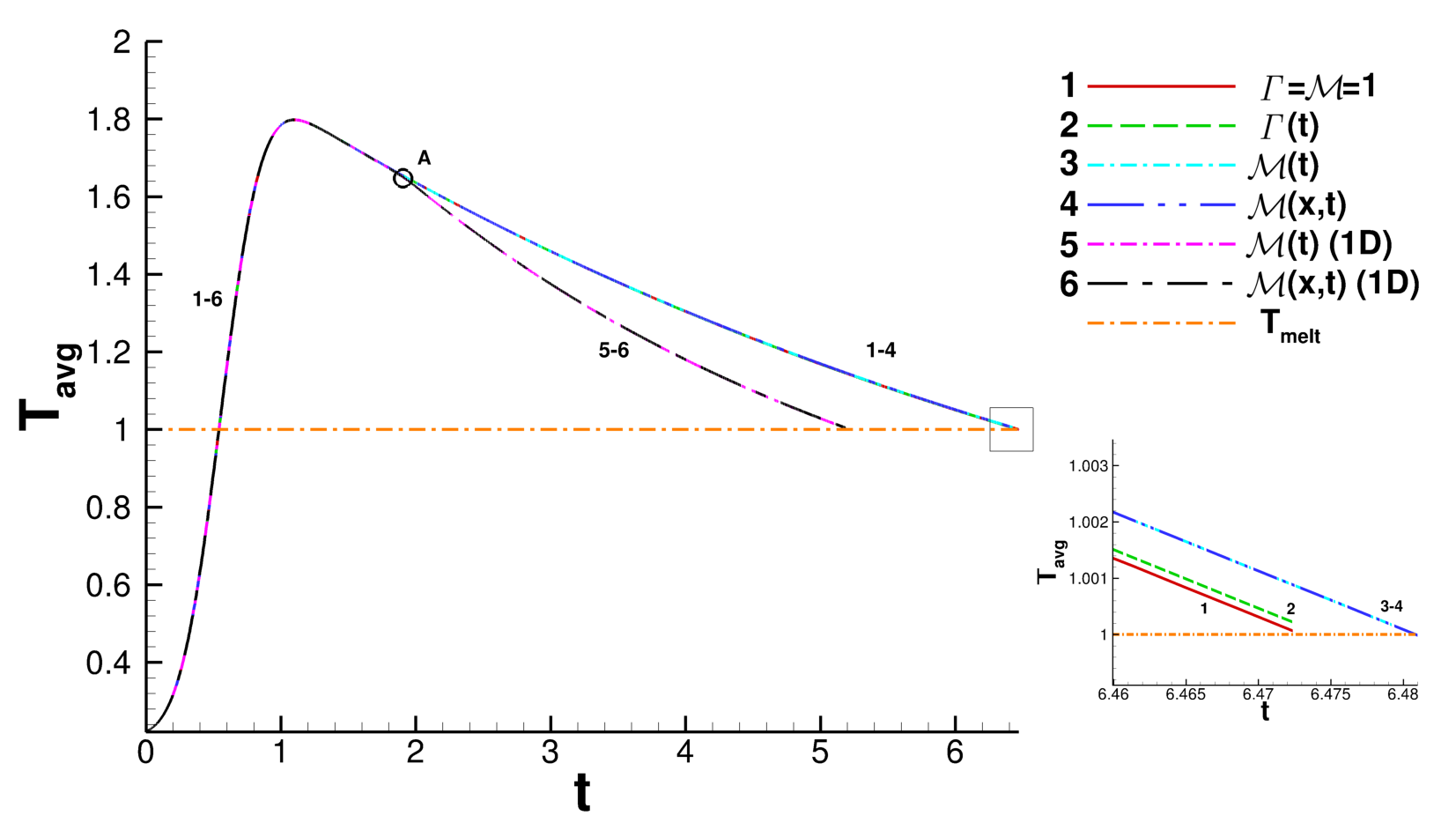}
    \caption{Average film temperatures with model (A): $\Gamma=1,\mathcal{M}=1$ ({\color{red} red} solid line); $\Gamma(t), \mathcal{M}=1$ ({\color{green} green} dashed line); $\Gamma=1, \mathcal{M}(t)$ ({\color{cyan} cyan} dash-dotted line); $\Gamma=1$, spatiotemporally-varying viscosity $\mathcal{M}(x,t)$ ({\color{blue} blue} dashed double-dotted line). Average film temperatures with model (1D): $\Gamma=1, \mathcal{M}(t)$, ({\color{magenta} magenta} dash-dotted line) and $\Gamma=1, \mathcal{M}(x,t)$, (black dash-dotted line). In all cases the domain length is $\Lambda_{\rm m}$ from LSA. Point $\textbf{A}$ marks the film dewetting time for the time-varying viscosity $\mathcal{M}(t)$ case. Inset: zoomed-in image of solidification  point for the model (A) cases. Here, including $\mathcal{M}(t)$ or $\mathcal{M}(x,t)$ leads to a slightly longer LL than the constant viscosity ($\mathcal{M}=1$) cases. Contrast with model (1D) where variable viscosity produces a LL that differs significantly from the rest. Simulations are marked $1$-$6$ and placed near the curves as a guide. $1$-$4$ were simulated with (A) and $5$-$6$ with (1D).}
    \label{fig:avg_temp_comparison}
\end{figure}
Figure~\ref{fig:avg_temp_comparison} plots the average film temperature profiles for each of the models; we observe that  
the temperatures agree up until point \textbf{A}, at which the film dewets for variable viscosity. The inset shows that, with model (A), the variable viscosity cases ($\mathcal{M}(t), \mathcal{M}(x,t)$) lead to a slightly longer LL than if viscosity is 
constant, $\mathcal{M}=1$. This may be important considering that the films dewet much nearer the resolidification time when $\mathcal{M}=1$. This difference, however, is always small relative to that between models (A) and (1D). When viscosity varies with time and/or space, model (1D) predicts LLs that are far shorter. Furthermore, for both models the results for $\mathcal{M}(x,t)$ and $\mathcal{M}(t)$ are identical. This is, we believe, due to the film temperatures not deviating far from the respective average temperatures (even for model (1D) where $x$-variation of temperature is much larger than for (A), the temperature deviates from the average by at most $30\%$, which appears to be insufficient to cause significant differences between results for $\mathcal{M}(t)$ and $\mathcal{M}(x,t)$). We conclude, therefore, that to simulate cases where film dewetting occurs much faster than resolidification, the dependence of viscosity on average temperature is the most important effect to include. 
 
\begin{figure}
    \centering
    \includegraphics[width=\textwidth]{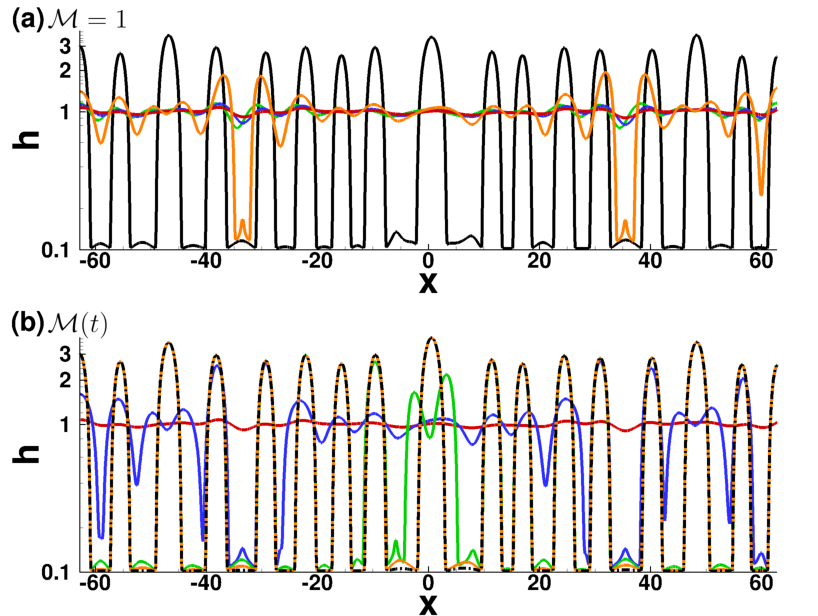}
    \caption{Evolution of films subject to initial random perturbations for (a) constant viscosity, $\mathcal{M}=1$, and (b) time-dependent viscosity, $\mathcal{M}(t)$. Color code: {\color{red} red} $t=0$; {\color{blue} blue} $t=1.66$; {\color{green} green} $t=2.30$; {\color{orange} orange} $t=3.31$;  black $t=6.47$ (re-solidification time for (a)). The domain length was taken to be $20\Lambda_{\rm m}$ and all simulations were done using model (A) with $\Gamma=1$. The $h$-axis is plotted on a log scale to emphasize satellite droplet formation.}
    \label{fig:thickness_rand_pert}
\end{figure}

To emphasize the importance of accurately modelling the film viscosity, we present simulation results for domains of length $20\Lambda_{\rm m}$, with films subjected to initial random perturbations according to Eq.~\eqref{h_random_pert}. Figure \ref{fig:thickness_rand_pert} plots film thickness for (a) $\mathcal{M}=1$ and (b) $\mathcal{M}(t)$, both using model (A) with $\Gamma=1$ and ${\rm Ma}=0$. We again see that films with time-dependent viscosity (Fig.~\ref{fig:thickness_rand_pert}b) dewet much faster than those with constant viscosity (Fig.~\ref{fig:thickness_rand_pert}a). The black curve represents the solidification time of (a) (nearly the same as that for (b)) and hence represents the final configurations of both simulations. The main finding is that the $\mathcal{M}=1$ case does not fully dewet, whereas the $\mathcal{M}(t)$ case does.
This difference in dewetting time scales is consistent with the earlier results for  domains of length $\Lambda_{\rm m}$, except that there the film was completely dewetted at solidification for both cases.  We note slight coarsening in Fig.~\ref{fig:thickness_rand_pert}a,b with some droplets merging, leading to fewer than 20 resultant (primary) droplets. 
 The film height evolution for model (1D), cases $\mathcal{M}(t), \mathcal{M}(x,t)$, and model (A) case $\mathcal{M}(x,t)$ is identical to Fig.~\ref{fig:thickness_rand_pert}b. 
 In the Appendix \ref{appendix_high_act_energy} we show a case where models (1D) and (A) lead to diverging results (for a different
 choice of parameters). We omit the temperature profiles here as the short LL for (1D) and identical $\mathcal{M}(t)$ and $\mathcal{M}(x,t)$ results are consistent with the results on domain length $\Lambda_{\rm m}$.

To summarize the results on long domains of length $20\Lambda_{\rm m}$, we have found that including the time-dependence of viscosity permits the films to fully dewet in the span of the liquid lifetime whereas keeping viscosity fixed at the melting temperature value ($\mathcal{M}=1)$ produces films that dewet only partially during the liquid phase. Any effects due to spatial variation of viscosity appear to be irrelevant here. Finally, the liquid lifetime is much shorter for model (1D) than model (A), as expected.



\section{Conclusions}
\label{conclusions}

To conclude, we have formulated three models for heat conduction in nanoscale thin films on thermally conductive substrates: a full model (F) that accounts for heat conduction in all directions in both film and substrate; an asymptotically-reduced model (A) that exploits a disparity in length scales in both film and substrate to derive an equation governing in-plane diffusion of heat within the film coupled to out-of-plane heat diffusion in the substrate; and a one-dimensional model (1D) model that simply neglects any in-plane diffusion in both film and substrate. In all cases a thin film model is used to describe the associated fluid dynamics. The main finding is that including in-plane diffusion in the thermal modeling influences strongly the film evolution. In particular, neglecting in-plane diffusion is found to amplify (artificially) in-plane thermal gradients and expedite film cooling. We have found that model (A) is 
significantly more accurate than (1D) while being considerably more computationally efficient than (F).  We have also found that when material parameters are allowed to vary in time through the average film temperature, model (A) produces liquid lifetimes significantly longer than those of model (1D), due to the absence of lateral heat conduction in (1D). Therefore model (A) combines both accuracy and efficiency.

With regard to the individual (dimensionless) material parameters that arise in our models, ${\rm Ma}$ (Marangoni number), $\Gamma$ (surface tension parameter)  and $\mathcal{M}$ (film viscosity), we find that the variation of viscosity with time has the greatest effect on model outcomes. By including time-dependent viscosity, films exposed to laser heating (on both small and large domains) fully dewet while in the molten state. In contrast, when viscosity is held constant, dewetting occurs much later in the cooling process, which may result in partial droplet formation only. This suggests strongly that time-dependent viscosity is needed to represent accurately experiment-like behavior. Using a spatiotemporally varying viscosity, ${\cal M}(x,t)$, produces essentially identical results to the case where viscosity depends only on time.
Introducing time dependence of the surface tension ($\Gamma(t)$) has a larger effect on the film instability growth rate (increasing it) than does the Marangoni effect (${\rm Ma}\neq 0$), though the effect is always small, and insignificant when compared to the variation of viscosity with time. The Marangoni effect was found to be negligible in all cases considered.

 Although model (A) is found to be useful in the current setting, its validity relies on a number of underlying assumptions. Therefore, its applicability to other problems must be carefully verified prior to use. In this study we have considered the time and space variation of only selected physical parameters (surface tension and viscosity) through temperature, and have not considered how temperature dependence of other material parameters, such as thermal conductivity and density, may influence the results. Furthermore, all simulations presented here are restricted to the two-dimensional geometry: much more significant computational benefit of model (A) is expected in three
 spatial dimensions (3D). Extending our model and simulations to 3D is one of the directions of our future work. 
 
\section*{Acknowledgement}
This research was supported by NSF DMS-1815613 (L.J.C., L.K.); by NSF DMS-1615719 (L.J.C., L.K.); by NSF CBET-1604351 (R.H.A., L.K.), and by CNMS2020-A-00110 (L.K.). The authors are grateful to Ivana Seric, William Batson, Michael Lam, Pejman Sanaei, and Shahriar Afkhami for fruitful discussions.

\section*{Declaration of Interests}
The authors report no conflict of interest.

\section*{Author ORCID}
R.H. Allaire, https://orcid.org/0000-0002-9336-3593, L.J. Cummings, https://orcid.org/0000-0002-7783-2126, and L. Kondic, https://orcid.org/0000-0001-6966-9851.


\appendix 

\section{Scalings and Parameter Values}

\subsection{Discussion of the choice of scales and Table of parameters}\label{appendix_scales}
 

\begin{table}
\centering
\setlength{\tabcolsep}{0.5em} 
{\renewcommand{\arraystretch}{1.17} 
  \begin{tabular}{ | l | l | l | l |} \hline 
\rowcolor{Gray} \textbf{Parameter} & \textbf{Notation} & \textbf{Value} & \textbf{Unit}  \\ \hline 
  Viscosity at Melting Temperature & $\prm{\mu}_{\rm f}$ (1) & $4.3 \times 10^{-3}$ & $\mathrm{Pa\cdot s}$  \\ \hline 
  Surface tension at Melting Temperature & $\prm{\gamma}_{\rm f}$ (1) & 1.303 & $\mathrm{J\cdot m^{-2}}$ \\ \hline
  Vertical length scale & $\prm{H}$ & $10$ & $\mathrm{nm}$  \\ \hline 
  Horizontal length scale & $\prm{L}=\prm{\lambda}_{\rm m}/(2\pi)$ & $40.58$ & $\mathrm{nm}$  \\ \hline 
  Time scale & $\prm{t}_{\rm s} = 3\prm{L}\prm{\mu}_{\rm f}/ (\epsilon^3 \prm{\gamma}_{\rm f})$ & $26.86$ & $\mathrm{ns}$ \\ \hline 
   Temperature scale/Melting Temperature & $\prm{T}_{\rm melt}$ & $1358$ & $\mathrm{K}$ \\ 
   \hline 
  Film density & $\prm{\rho}_{\rm f}$ (1) & $8000$ & $\mathrm{kg \cdot m^{-3}}$  \\ \hline 
  SiO$_2$ density & $\prm{\rho}_{\rm s}$ (1)  & $2200$ & $\mathrm{kg \cdot m^{-3}}$  \\ \hline
  Film specific heat capacity & $\prm{c}_{\rm f}$ (1) & $495$ & $\mathrm{J \cdot kg^{-1} \cdot K^{-1}}$ \\ \hline 
  SiO$_2$ specific heat capacity & $\prm{c}_{\rm s}$ (1) & $937$ & $\mathrm{J \cdot kg^{-1} \cdot K^{-1}}$ \\ \hline 
  Film heat conductivity & $\prm{k}_{\rm f}$ (1) & $340$ & $\mathrm{W \cdot m^{-1} \cdot K^{-1}}$  \\ \hline
    SiO$_2$ heat conductivity & $\prm{k}_{\rm s}$ (1) & $1.4$ & $\mathrm{W \cdot m^{-1} \cdot K^{-1}}$\\ \hline 
  Film absorption length & $\prm{\alpha}_{\rm f}^{-1} H$ (1) & $11.09$ & $\mathrm{nm}$  \\ \hline
  Temp. Coeff. of Surf. Tens. & $ \prm{\gamma}_{\rm T}$ (1) & $-0.23 \times 10^{-3}$ & $\mathrm{J \cdot m^{-2} \cdot K^{-1}}$  \\ \hline 
    Hamaker constant & $\prm{A}_{\rm H}$ (2) & $1.75\times 10^{-17}$ & $\mathrm{J}$  \\ \hline
  Reflective coefficient & $r_0$ (1) & $0.3655$ & 1  \\ \hline
  Film reflective length & $\prm{\alpha}_{\rm r}^{-1} H$ (1) & $12.0$ & $\mathrm{nm}$ \\ \hline 
    Laser energy density & $\prm{E}_0$ (3) & $300$ & $\mathrm{J \cdot m^{-2}}$  \\ \hline
    Gaussian pulse peak time & $\prm{t}_{\rm p} t_{\rm s}$ (3) & $15$ & $\mathrm{ns}$ \\ \hline
    Equilibrium film thickness & $\prm{h}_* H$ & $1$ & $\mathrm{nm}$  \\ \hline 
  Mean Film thickness & $\prm{h}_0 H$ & $10$ & $\mathrm{nm}$ \\ \hline 
  SiO$_2$ thickness & $\prm{H}_{\rm s} H$ & $10$ & $\mathrm{nm}$  \\ \hline
  Room temperature & $\prm{T}_{\rm a} T_{\rm melt}$ & $300$ & $\mathrm{K}$ \\ \hline 
  SiO$_2$ Heat Transfer Coefficient & $\prm{\alpha}$ & $3.0 \times 10^{5}$ & $\mathrm{W \cdot m^{-2} \cdot K^{-1}}$ \\ \hline
    Characteristic Velocity & $\prm{U}$ & $1.504$ & $\mathrm{m \cdot s^{-1}}$ \\ \hline
    Activation Energy & $E$ & $30.5$ & ${\rm kJ} \cdot {\rm mol}^{-1}$ \\ \hline
    \rowcolor{Gray} \textbf{Non-Dimensional Numbers} &  & & \textbf{Expression} \\ \hline 
     Aspect Ratio & $\epsilon$ & $0.246$ & $H/L$ \\ \hline 
    Reynolds Number & $\mbox{Re}$ & $0.114$ & $\rho_{\rm f} U L/ \mu_{\rm f}$ \\ \hline
    Film Peclet Number & $\mbox{Pe}_{\rm f}$ & $7.14 \times 10^{-4}$ & $(\rho c)_{\rm f} U L/k_{\rm f}$ \\ \hline
    Substrate Peclet Number & $\mbox{Pe}_{\rm s}$ & $5.46 \times 10^{-3}$ & $(\rho c)_{\rm s} U \epsilon H/k_{\rm s}$ \\ \hline
    Biot Number & ${\rm Bi}$ & $2.14 \times 10^{-3}$ & $\alpha H/k_{\rm s}$ \\ \hline 
    Marangoni Number & ${\rm Ma}$ & $0.360$ & $3\gamma_{\rm T}T_{\rm melt}/(2\gamma_{\rm f})$ \\  \hline
   Thermal Conductivity Ratio & $\mathcal{K}$ & $0.068$ & $k_{\rm s}/(\epsilon^2 k_{\rm f})$ \\ \hline
   Range of Dimensionless Viscosity & $\mathcal{M}$ & $0.028-1$ & $\mu/\mu_{\rm f}$ \\ \hline
 \end{tabular} }
 \caption{Parameters used for liquid Cu film and SiO$_2$ substrate. References: (1) \citet{Dong_prf16}, (2) \citet{lang13}, (3) \cite{mckeown12}.}
 \label{table:ref_paras}
\end{table}

The choice of scales has important implications for the derivation of both the thin film equation (Eq. \eqref{thin_film}) and the thermal model (A) (Eq. \eqref{asymptotic model1}-\eqref{asymptotic_model_end}). The choice of timescale is typically based on the fluid flow, $L/U$. From the perspective of the thermal model, however, the pulsed laser heating duration is on the order of nanoseconds. With the timescale choice $L/U$, $L$ and $U$ should be chosen consistent with such a thermal time scale, while still retaining surface tension effects in the fluid flow model, known to be important. Therefore, we choose $U=\epsilon^3 \gamma_{\rm m}/(3\mu_{\rm m})$ and scale  $L$ on the (inverse) wavenumber of maximum growth $k_{\rm m}= 2\pi/\lambda_{\rm m}$ as $L=k_{\rm m}^{-1}$ (from  \S\ref{lsa_section}) so that surface tension effects and disjoining pressure are retained to leading order. The tradeoff is that the aspect ratio $\epsilon$ consistent with the data (Table~\ref{table:ref_paras}) is rather large, 0.246. However, the time scale is then on the order of nanoseconds, as desired, and we consider this acceptable in order to develop a consistent model.

For the materials in question, the values of ${\rm Pe}_{\rm f}, {\rm Pe}_{\rm s}, \mathcal{K}, $ and ${\rm Bi}$ are small despite the $O(1)$ assumption (Table~\ref{table:ref_paras}). This is primarily a consequence of the size of $\epsilon$, which is directly related to the dependence of $L$ on $\lambda_{\rm m}$. Given this observation, we briefly consider the limit of small ${\rm Pe}_{\rm f}, {\rm Pe}_{\rm s}, \mathcal{K},$ and ${\rm Bi}$. Firstly, in the limit ${\rm Pe}_{\rm f}\to 0$, Eq.~\eqref{asymptotic model1} would reduce to a quasi-steady (no time-derivative) equation governing temperature $T_{\rm f}(x,y,t)$ in the film. The resultant equation is computationally more difficult to solve (and leads to only negligible differences), so we do not adopt this approach. Secondly, in the limit ${\rm Pe}_{\rm s} \to 0$, the solution to Eq.~\eqref{asymptotic_sub_eqn} would be linear in $z$. The numerical solutions given in Fig. \ref{fig:Tfilm_and_Tsub_full} display substrate temperatures that deviate from  linear behaviour in $z$. We consider this the motivation for retaining ${\rm Pe}_{\rm s}$ in Eq.~\eqref{asymptotic_sub_eqn} (which leads to better agreement between models (A) and (F)). In the limit $\mathcal{K}\to 0$, film temperature no longer depends directly on substrate temperature (although the substrate temperature still depends on the film temperature). Since the primary heat loss mechanism is considered to be through the film-substrate interface, we retain $\mathcal{K}$ in Eq.~\eqref{asymptotic model1}. If the primary heat loss mechanism is elsewhere it may be possible to drop the term containing $\mathcal{K}$ in Eq.~\eqref{asymptotic model1}. Finally, in the limit ${\rm Bi} \to 0$, Eq.~\eqref{Bi_boundary_condition} becomes an insulating boundary condition, which in turn leads to much higher substrate/film temperatures. In the case where SiO$_2$ sits on a native layer of Si it is expected that there is some heat transfer. Therefore, we retain ${\rm Bi}$ in Eq.~\eqref{Bi_boundary_condition}.  

\subsection{Wavelength of Maximum Growth}\label{appendix_lsa}
When the material parameters are fixed at melting temperature, $\Gamma,\mathcal{M}=1$, the dimensional dispersion relation can be written as:
\begin{align}
    \beta(k) &= \frac{h_0^2 H^2 k^2}{3\mu_{\rm f}} \left( P_0 - \gamma_{\rm f} H h_0 k^2 \right), \\
    P_0 &= \frac{A_{\rm H}}{6\pi h_*^3 H^3} \left[ m\left(\frac{h_*}{h_0} \right)^m - n \left( \frac{h_*}{h_0}\right)^n \right],
\end{align}
where (in this section only) $\beta(k)$ is the dimensional growth rate and $k$ the dimensional wavenumber. The remaining parameters are given in Table \ref{table:ref_paras}. The wavelength of maximum growth, $\lambda_{\rm m}$, can then be found by setting $\partial \beta/\partial k=0$ and written as
\begin{align}
    \lambda_{\rm m} = 2\pi\bigg/ \sqrt{\frac{A_{\rm H}}{12\pi \gamma_{\rm f} h_*^3 H^3 h_0}\left[ m \left( \frac{h_*}{h_0} \right)^m - n \left( \frac{h_*}{h_0}\right)^n \right]}. \nonumber
\end{align}
For the parameters used in this paper $\lambda_{\rm m}=255{\rm nm}$. For the entirety of the paper, this value is used to define a base length scale $L=\lambda_{\rm m}/(2\pi)$.

\section{Effect of spatially varying viscosity with a larger Biot number}\label{appendix_high_act_energy}
Here we increase the Biot number to ${\rm Bi}=5.71\times 10^{-3}$ (more than twice the value used in the main text; see Table~\ref{table:ref_paras}). This leads to much earlier resolidification of the film (for (A) resolidification occurs at $t\approx 2.8$ whereas for (1D) $t\approx 2.7$). We specifically focus on the influence of model choice when the spatiotemporally-varying viscosity is used, $\mathcal{M}(x,t)$.
\begin{figure}[H]
    \centering
    \includegraphics[width=\textwidth]{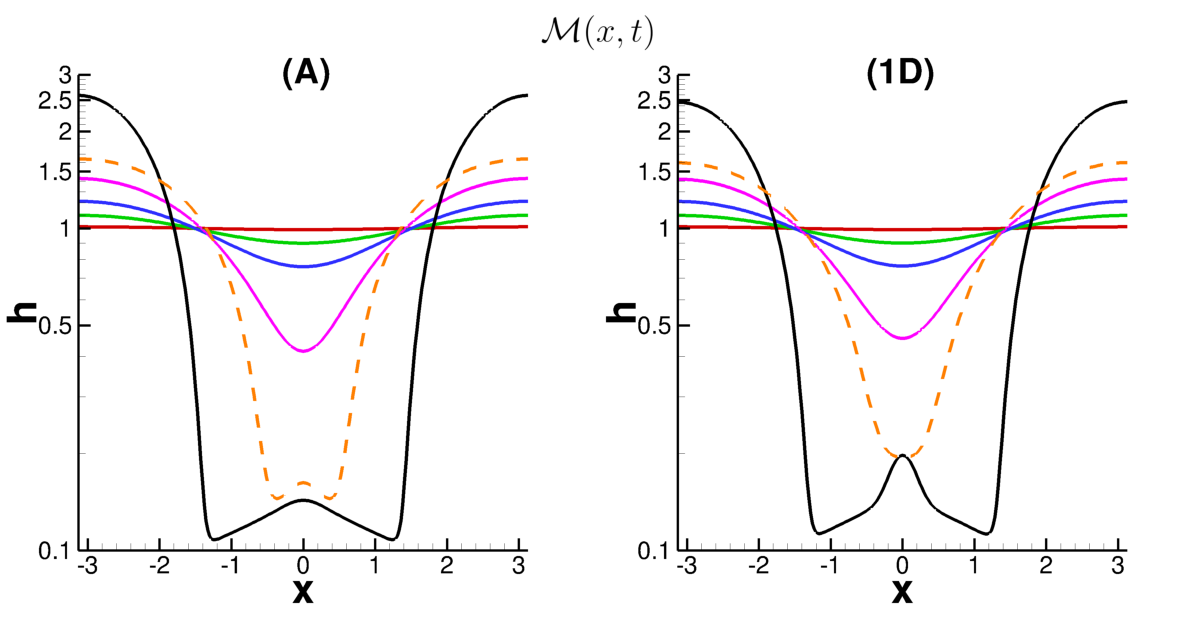}
    \caption{Film thickness evolution for both the asymptotic model (left) and the 1D model (right) for the spatiotemporal varying viscosity case $\mathcal{M}(x,t)$. The $y$-axis is plotted on a log scale to emphasize satellite droplet formation which is more prominent with (1D) than with (A). Here the Biot number is ${\rm Bi}=5.71\times 10^{-3}$. The times are: $t=0$ ({\color{red} red}),  $t=1.47$ ({\color{green} green}), $t=1.84$ ({\color{blue} blue}), $t=2.21$ ({\color{magenta} magenta}), $t=2.40$ ({\color{orange} orange} dashed), resolidification of (1D), $t=2.69$ (black).}
    \label{fig:appendix_heat_transfer_coef}
\end{figure}
Figure \ref{fig:appendix_heat_transfer_coef} shows the film evolution for both models (A) (left) and (1D) (right). The black solid line represents the film profiles at the (1D) solidification time (model (A) predicts a larger solidification time but the final solidified film configuration is nearly identical to the black solid line in Fig.~\ref{fig:appendix_heat_transfer_coef}a). The times are given in the caption. The main finding is that for sufficiently large $\alpha$ the decay of the satellite droplet that forms for both (A) and (1D) is slower in (1D). Essentially, for (1D) the satellite droplet is cold relative to the thicker parts of the film (recall the much more significant $x$-variation of temperature observed for (1D)). Since the viscosity varies with space the centre ($x=0$) of the satellite droplet approaches the precursor thickness more slowly than the surrounding area. In (A), the temperature variation with $x$ is less pronounced and so the satellite droplet has drained more than its (1D) counterpart before reaching the final solidified configuration. In summary, these results demonstrate that the choice of thermal model can influence the final resulting film profiles. 
\section{Numerical Schemes}

\subsection{Numerical Solution of model (F)} \label{numerical_F_section}
To solve equation \eqref{system5} numerically, along with corresponding boundary conditions in model (F), we define the new variables $(u,v,\tau)$ as:
\begin{align}
    u=x, \qquad v= \frac{zh_0}{h}, \qquad \tau=t, \nonumber
\end{align}
which is a time-dependent mapping transforming the deformable domain, describing the film, into a fixed rectangle (see Fig. \ref{fig:mapping_schematic} below).
\begin{figure}[H]
    \centering
    \includegraphics[width=\textwidth]{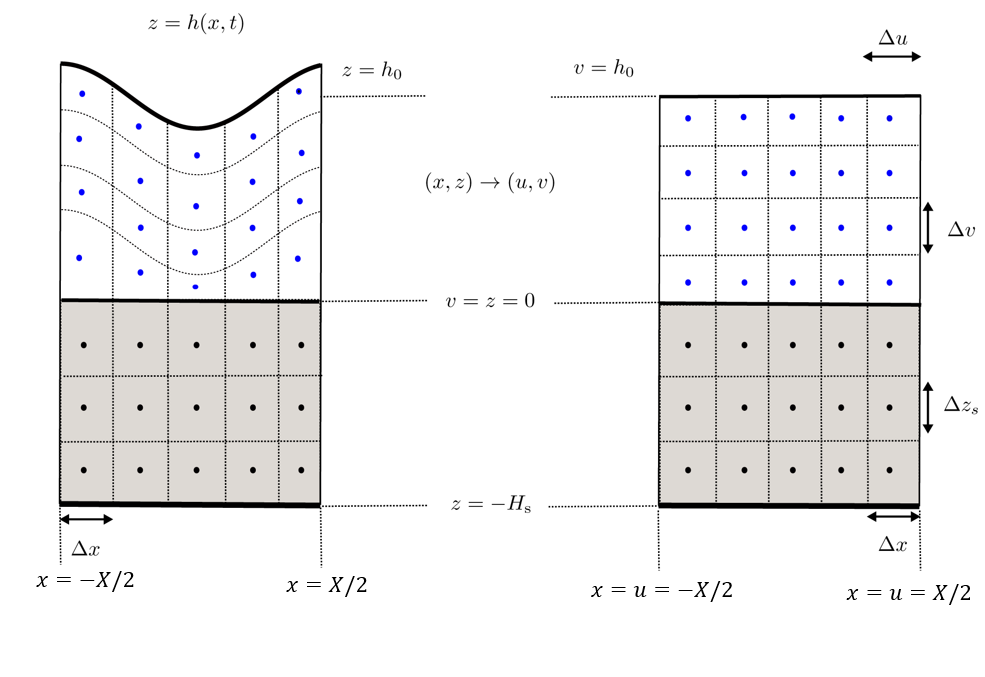}
    \caption{Schematic of mapping used to transform the computational domain for (F). Here the domain width shown is $X=2\pi$. Blue points represent the film domain and black points represent the substrate domain.}
    \label{fig:mapping_schematic}
\end{figure}

This approach trades a simplified domain for an increase in complexity of the thermal equation in the film. Equation \eqref{system5} is transformed into the following differential equation:
\begin{align}
 \partial_{\tau} T_{\rm f} = \frac{1}{{\rm Pe}_{\rm f}} \partial_u^2 T_{\rm f} + B(u,v) \partial_{uv} T_{\rm f} + A(u,v) \partial_v^2 T_{\rm f} +  F(u,v) \partial_v T_{\rm f} + Q(h(u),v,\tau), \label{heat_trans}
\end{align}

where
\begin{align}
B(u,v)&= -\frac{1}{{\rm Pe}_{\rm f}}\left( \frac{2v \partial_u h}{h} \right), \nonumber \\
A(u,v)&= \frac{1}{{\rm Pe}_{\rm f}} \left[ \left( \frac{v\partial_u h}{h} \right)^2 + \frac{1}{\epsilon^2} \left( \frac{h_0}{h} \right)^2 \right], \nonumber \\
F(u,v)&= \frac{1}{{\rm Pe}_{\rm f}}\left( \frac{vh\partial_u^2 h}{h^2} - \frac{v (\partial_u h)^2}{h^2} \right) + \frac{v \partial_{\tau} h}{h}, \nonumber \\
Q(h(u),v,\tau) &= \frac{F(t)}{{\rm Pe}_{\rm f}} \left[ 1-R(h) \right] \exp{\left[-\alpha_{\rm f}\left(h-hv/h_0 \right)\right]}, \nonumber
\end{align}
are the coefficients, and subscripts $u,v,\tau$ denote partial derivatives. 
To solve equation \eqref{heat_trans} numerically the ADI method is used, with the term containing $B(u,v)$ treated explicitly. A Crank-Nicolson scheme is used to solve Eq. \eqref{system6}, the heat equation in the substrate. We use a cell-centered grid system:
\begin{align} 
    u_i = x_i &= x_0 + \Delta x \left( i - 1/2 \right), \quad i=1,\ldots, n, \qquad
    \Delta u = \Delta x = \frac{\left(x_{\rm max} - x_0 \right)}{n}, \nonumber \\
    z_j &= -\left(j-1/2 \right) \Delta z_s, \quad j=1,\ldots, n_{\rm s}, \qquad
    \Delta z_s = \frac{H_{\rm s}}{n_{\rm s} }, \nonumber \\
    v_j &= h_0 - \left( j-1 \right) \Delta v, \quad j=1,\ldots, n_{\rm f}, \qquad
    \Delta v = \frac{h_0}{n_{\rm f}}, \nonumber \\
    T_{i,j}^k &\approx \begin{cases} T_{\rm f}(u_i, v_j, t_k), \qquad 1<j<n_{\rm f}, \\ T_{\rm s}(x_i, z_j, t_k), \qquad n_{\rm f}+1 < j < n_{\rm f} + n_{\rm s},    \end{cases} \nonumber \\
        h_i^k &\approx h(u_i, t_k), \quad Q_{i,j}^k \approx  Q_(u_i, v_j, t_k). \nonumber
\end{align}
The numerical system can be then written as:

\begin{align}
    \frac{T_{i,j}^* - T_{i,j}^k}{\Delta t} &= D_{i,j}^* + G_{i,j}^k + M_{i,j}^k + \frac{1}{2} Q_{i,j}^{k+1/2}, \nonumber  \\
    \frac{T_{i,j}^{k+1} - T_{i,j}^*}{\Delta t} &= D_{i,j}^* + G_{i,j}^{k+1} + \frac{1}{2} Q_{i,j}^{k+1/2}, \nonumber \\
    D_{i,j}^* &= \begin{cases} \frac{1}{2}{\rm Pe}_{\rm f}^{-1} \delta_u^2 T_{i,j}^*, &j\leq n_{\rm f}, \\
    \frac{1}{2} \epsilon^2 {\rm Pe}_{\rm s}^{-1} \delta_x^2 T_{i,j}^*, &j\geq n_{\rm f}+1, \end{cases} \nonumber \\
    G_{i,j}^k &= \begin{cases} \frac{1}{2} A_{i,j}^k \delta_v^2 T_{i,j}^k + \frac{1}{2} F_{i,j}^k \delta_v T_{i,j}^k, &j\leq n_{\rm f}, \\
    \frac{1}{2} {\rm Pe}_{\rm s}^{-1}\delta_z^2 T_{i,j}^k, &j\geq n_{\rm f}+1, \end{cases} \nonumber \\
    M_{i,j}^k &= \begin{cases} B_{i,j}^k \delta_{uv} T_{i,j}^k, & j\leq n_{\rm f}, \\
    0, & j>n_{\rm f}, 
    \end{cases} \nonumber
\end{align}
where $T_{\rm f}(x_i,z_j,t_k)\approx T_{i,j}^{k}$ is a discretisation of the film temperature, and $T_{i,j}^{*}$ represents the solution at an intermediate time step. In the interior grid the spatial derivatives are given by:
\begin{align}
\delta_u T_{i,j}=\frac{T_{i+1,j}-T_{i-1,j}}{2\Delta u}, \qquad \delta_v T_{i,j}=\frac{T_{i,j-1}-T_{i,j+1}}{2\Delta v}, \nonumber \\
\delta_x^2 T_{i,j} = \delta_u^2 T_{i,j}= \frac{T_{i+1,j}-2T_{i,j}+T_{i-1,j}}{\Delta u^2}, \nonumber \\
\delta_{uv} T_{i,j} = \frac{T_{i+1,j-1}-T_{i-1,j-1}-T_{i+1,j+1}+T_{i-1,j+1}}{4\Delta u \Delta v}, \nonumber \\
\delta_z^2 T_{i,j} = \frac{T_{i,j-1}-2T_{i,j} + T_{i,j+1}}{\Delta z_s^2}, \qquad \delta_v^2 T_{i,j}=\frac{T_{i,j+1}-2T_{i,j}+T_{i,j-1}}{\Delta v^2}, \nonumber
\end{align}
which are second-order central difference approximations and the source, $Q$, is approximated by an average at times $t_k$, and $t_{k+1}$, 
\begin{align}
    Q_{i,j}^{k+1/2} &= \frac{1}{2} \left( Q_{i,j}^{k+1} + Q_{i,j}^k \right). \nonumber
\end{align}

\subsection{Numerical Solution of model (A)}\label{numerical_A_section}

We use the same cell-centered grid as \S\ref{numerical_F_section} except for $z_j$:
\begin{align}
        z_j &= -\left(j-1 \right) \Delta z_{\rm s}, \quad j=1,\ldots, n_{\rm s}, \qquad
    \Delta z_{\rm s} = \frac{H_{\rm s}}{n_{\rm s} - 1/2}, \nonumber
\end{align}
so that a grid point exists on the liquid-solid interface (a Dirichlet boundary condition is prescribed there). For simplicity we let $T_i^{k} \approx T_{\rm f}(x_i, t_k)$, $S_{i,j}^{k} \approx T_{\rm s}(x_i,z_j,t_k)$ and

\begin{align}
    F_i^{k} &= \frac{1}{{\rm Pe}_{\rm f}} \left[ \delta_x^2 T_i^{k} + \left( \frac{\partial_x h}{h}\right)_{i}^{k} \delta_x T_{i}^{k} \right], \nonumber \\
    G_i^{k} &= -\frac{\mathcal{K}}{ {\rm Pe}_{\rm f} h_i^k} \partial_z \left( S \right)_I^{k} ,  \nonumber  
\end{align}
where $\partial_z (S)_{I}^k=\partial T_{\rm s}/\partial z \vert_{z=0}(t=t_k)$. To solve Eq. \eqref{asymptotic model1} and \eqref{asymptotic_sub_eqn} we use a predictor-corrector Runge-Kutta/Crank-Nicolson scheme. In the prediction phase we use a forward-Euler scheme to deal with $G_i^k$:
\begin{align}
\frac{T_{i}^{k+1}-T_{i}^k}{\Delta t}&=\frac{1}{2} \left[ F_i^{k+1} + F_i^{k} \right]  + \frac{1}{2} \left( G_i^{k} + \hat{G}_i^{k} \right) + \overline{Q}_{i}^{k+1/2}, \quad i= 1, \ldots, n, \nonumber \\
\frac{S_{i,j}^{k+1} - S_{i,j}^{k}}{\Delta t} &= \frac{1 }{2} {\rm Pe}_{\rm s}^{-1} \left[ \delta_z^2 S_{i,j}^{k+1} + \delta_z^2 S_{i,j}^{k} \right], \quad j= 2, \ldots, n_{\rm s}, \qquad i = 1, \ldots, n,  \nonumber
\end{align}
where $\overline{Q}_i^{k+1/2}=(\overline{Q}_i^k+\overline{Q}_i^{k+1})/2$. We then correct this prediction by using a Runge-Kutta, order 2, method on $G_i^k$ using the prediction $\hat{G}_i^k$:
\begin{align}
\frac{T_{i}^{k+1}-T_{i}^k}{\Delta t}&=\frac{1}{2} \left[ F_i^{k+1} + F_i^{k} \right]  + \frac{1}{2} \left( G_i^{k} + \hat{G}_i^{k} \right) + \overline{Q}_{i}^{k+1/2}, \quad i = 1, \ldots, n, \nonumber \\
\frac{S_{i,j}^{k+1} - S_{i,j}^{k}}{\Delta t} &= \frac{1 }{2} {\rm Pe}_{\rm s}^{-1} \left[ \delta_z^2 S_{i,j}^{k+1} + \delta_z^2 S_{i,j}^{k} \right], \quad j = 2, \ldots, n_{\rm s}, \qquad i = 1, \ldots, n. \nonumber
\end{align}

\subsection{Numerical Solution of model (1D)}
The numerical scheme used for model (1D) is a simple Crank-Nicolson scheme:
\begin{align}
    \frac{T_{i,j}^{k+1} - T_{i,j}^k}{\Delta t} &= \frac{1}{2}\epsilon^{-2}{\rm Pe}_{\rm f}^{-1} \left( \delta_z^2 T_{i,j}^k + \delta_z^2 T_{i,j}^{k+1} \right) + Q_{i,j}^{k+1/2}, \qquad &&j=1,\ldots,n_{\rm f}  \nonumber \\
    \frac{T_{i,j}^{k+1} - T_{i,j}^k}{\Delta t} &= \frac{1}{2}{\rm Pe}_{\rm s}^{-1} \left( \delta_z^2 T_{i,j}^k + \delta_z^2 T_{i,j}^{k+1} \right), \qquad &&j=n_{\rm f}+1,\ldots, n_{\rm f}+n_{\rm s}, \nonumber
\end{align}
where $i=1,\ldots, n$ and $n, n_{\rm f}$ and $n_{\rm s}$ are the same as in \S\ref{numerical_F_section}.

\subsection{Convergence Results}
For what follows we define the discrete $L2$ error, $E(t)$, where
\begin{align}
    E^2(t) = \frac{\sum\limits_{i=1}^{n} \left(\sum\limits_{j=1}^{n_{\rm f}}\Delta z_i \left| T^{\rm comp}_{i,j}-T^{\rm bench}_{i,j} \right|^2 + \Delta z_{\rm s} \sum\limits_{j=1}^{n_{\rm s}} \left| T^{\rm comp}_{i,j}-T^{\rm bench}_{i,j} \right|^2 \right) }{\sum\limits_{i=1}^{n} \left(\sum\limits_{j=1}^{n_{\rm f}}\Delta z_i \left|T^{\rm bench}_{i,j} \right|^2 + \Delta z_{\rm s} \sum\limits_{j=1}^{n_{\rm s}} \left| T^{\rm bench}_{i,j} \right|^2 \right) }, \nonumber
\end{align}
and $T_{i,j}^{\rm comp}$ is the computed temperature and $T_{i,j}^{\rm bench}$ is a benchmark solution which, for the results presented next, we take to be the numerical solution on the finest grid since no exact solution is known.

\begin{figure}[H]
    \centering
    \includegraphics[width=\textwidth]{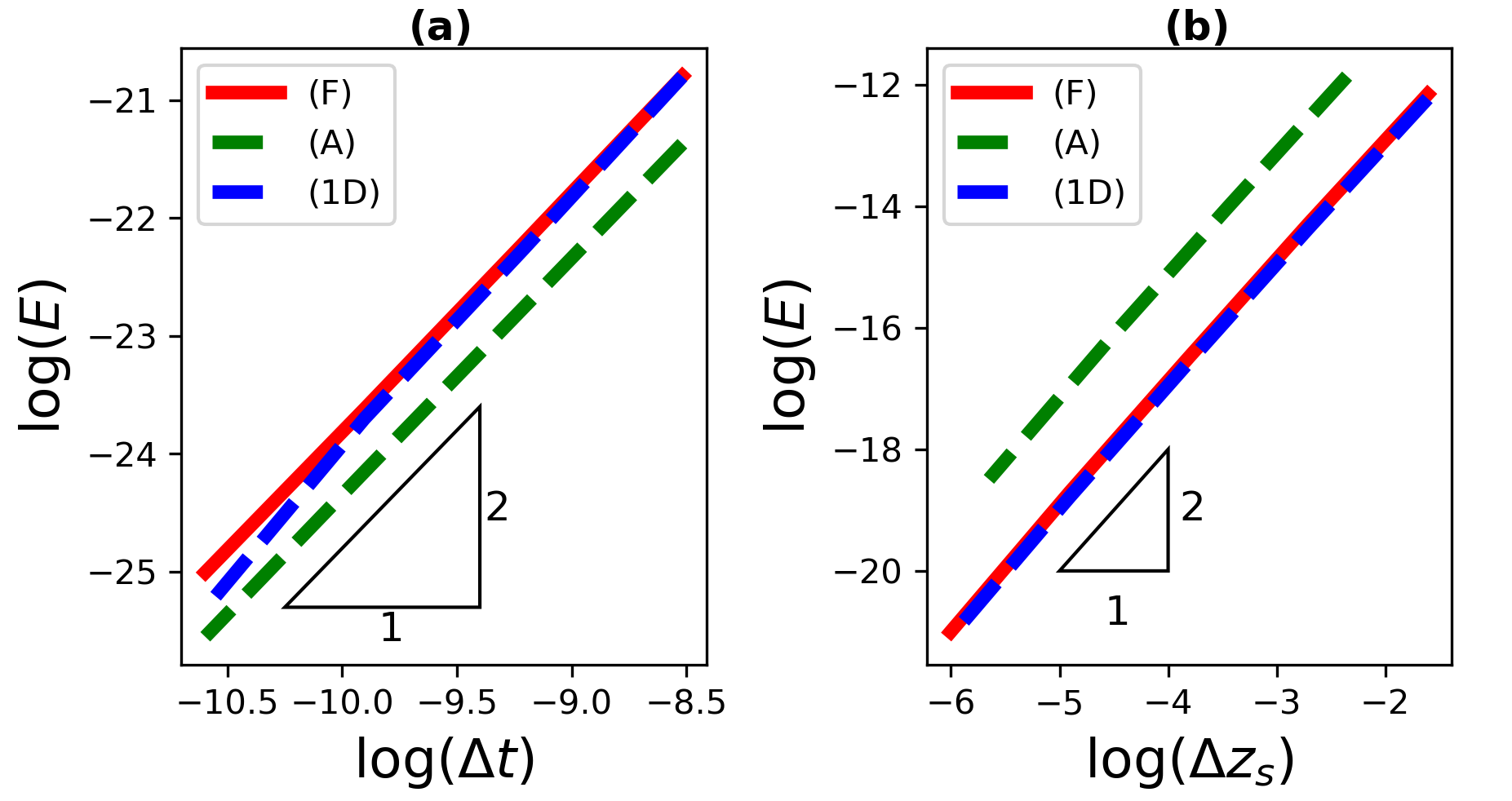}
    \caption{\textbf{(a)}$\Delta t$ convergence for (F), (A), and (1D) in the stationary curved film case. All models use $O(\Delta t^2)$ schemes. \textbf{(b)}$\Delta z_{\rm s}$ convergence for (F), (A), and (1D) in the stationary curved film case where $h$ is given by Eq.~\eqref{h_ic1}, but time-independent. All models use $O(\Delta z_{\rm s}^2)$ schemes.}
    \label{fig:dt_and_dzs_conv}
\end{figure}

\begin{figure}[H]
    \centering
    \includegraphics[width=\textwidth]{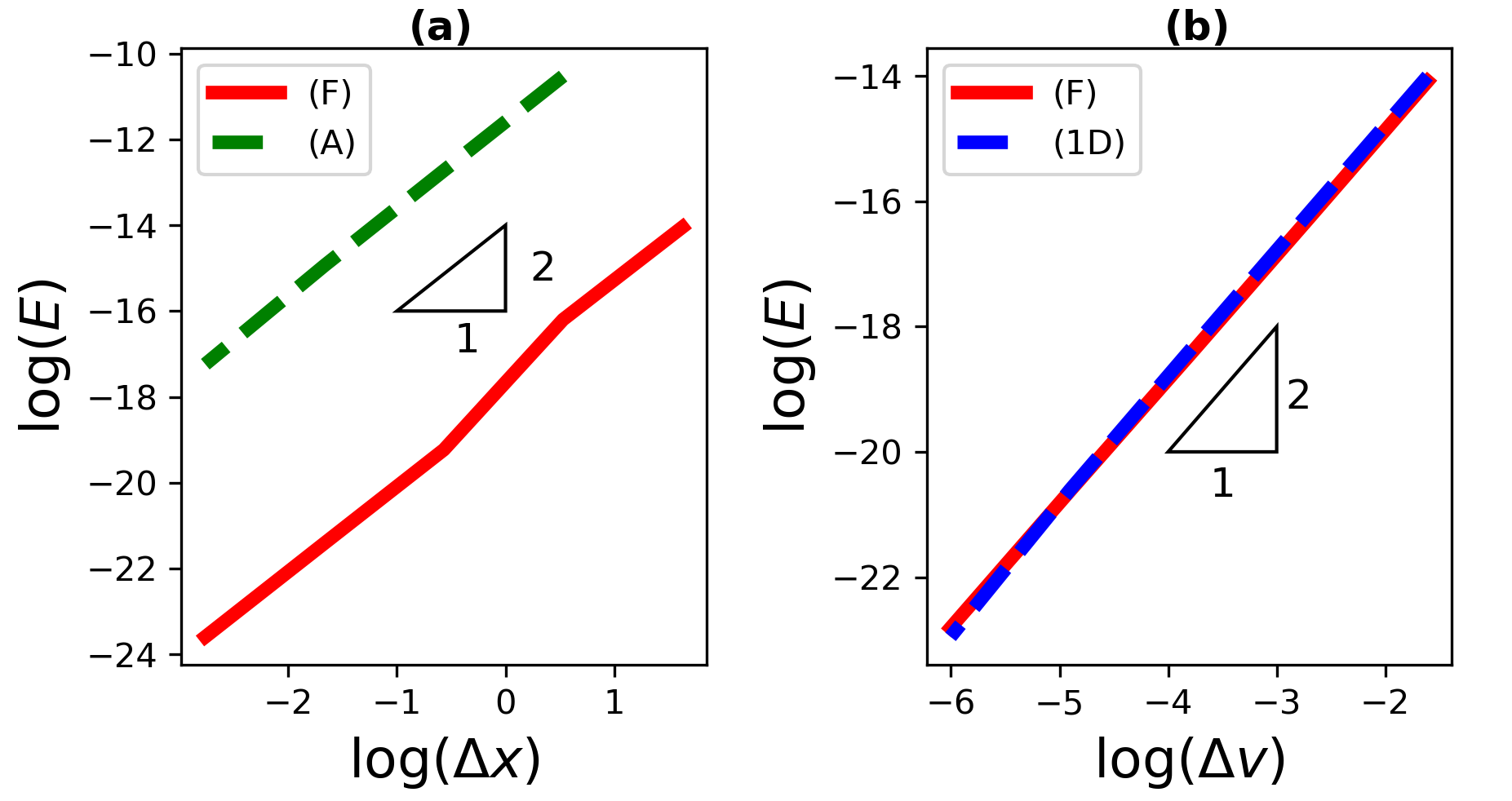}
    \caption{\textbf{(a)}$\Delta x$ convergence for (F) and (A) in the stationary curved film case (Note that (1D) has no derivatives with respect to $x$). Each model uses a $O(\Delta x^2)$ scheme. \textbf{(b)}$\Delta v$ convergence for (F) and (1D) in the stationary curved film case (Note that $T_{\rm f}$ is independent of $z$ for (A)). Each model uses a $O(\Delta v^2)$ scheme.}
    \label{fig:dx_and_dv_conv}
\end{figure}

\section{Variation of Temperature with Film Thickness}\label{appendix_dTdh}

In the case when the film is flat its temperature is independent of the in-plane variables and conservation of energy may be reduced to an expression for the average film temperature, written as a simple balance of source heating and substrate cooling:
\begin{align}
    \partial_t T_{\rm avg} = {\rm Pe}_{\rm f}^{-1} \left[ \frac{-\mathcal{K}}{h} \left( \partial_z T_{\rm s}\right)\vert_{z=0} + \overline{Q} \right].
\end{align}
\begin{figure}[H]
    \centering
    \includegraphics[width=\textwidth]{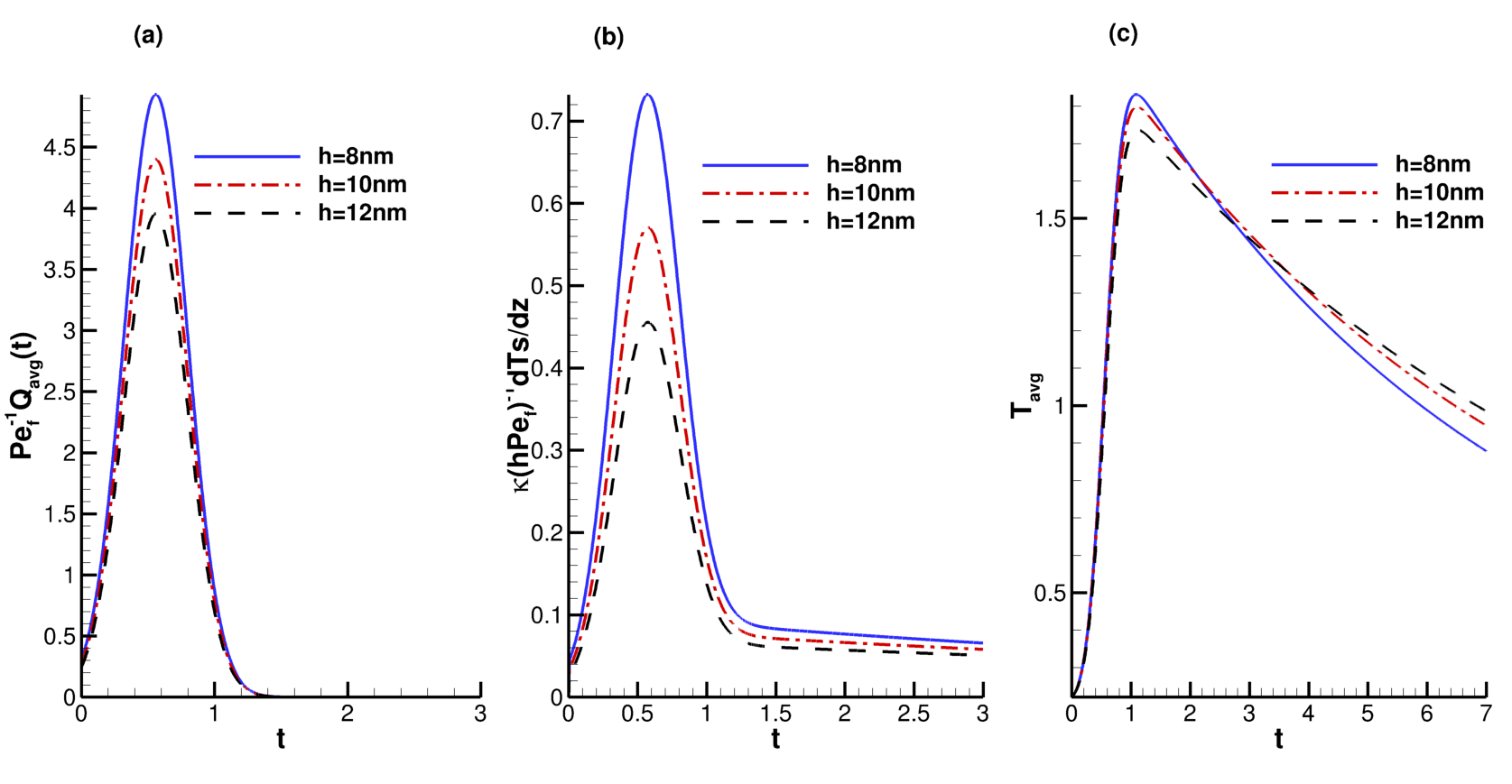}
    \caption{(a) Average Source, $\overline{Q}$, and (b) magnitude of film heat loss through the substrate, $\mathcal{K}(h {Pe}_{\rm f})^{-1}\partial_z T_{\rm s}\vert_{z=0}$ and (c) average film temperature all for flat films of $h=8$nm ({\color{blue} blue}, solid), $h=10$nm ({\color{red} red}, dot-dashed), and $h=12$nm (black, dashed).}
    \label{fig:flat_film_heating_cooling}
\end{figure}

Figure \ref{fig:flat_film_heating_cooling} shows the variation of both (a) source heating (measured by ${\rm Pe}_{\rm f}^{-1}\overline{Q}$) and (b) substrate cooling (measured by $\left( h {\rm Pe}_{\rm f} \right)^{-1} \left( \partial_z T_{\rm f}\right)\vert_{z=0}$) for three different film thicknesses, $h=8,10,12$nm. Figure \ref{fig:flat_film_heating_cooling}(c) gives the average film temperatures for each case. As seen in Fig.~\ref{fig:flat_film_heating_cooling}(a) thinner films retain more energy from the source and, in the absence of cooling, should be hotter. Physically, thicker films reflect more energy and absorb less. Although (to keep the presentation simple) we use $\overline{Q}$ here rather than $Q$ from Eq.~\eqref{Q_eqn}, it can be shown also that $dQ/dh<0$ for all times. Figure \ref{fig:flat_film_heating_cooling}(b) demonstrates that thinner films also cool faster (through the substrate) than thicker ones. When these two effects are combined, we arrive at average film temperatures that are non-monotonic in film thickness $h$. In Fig.~\ref{fig:flat_film_heating_cooling}(c) thinner films are observed to be initially hotter over the early stages of evolution, primarily due the magnitude of the laser heating, which is initially larger than that of cooling. Over the later stages, the source decreases in strength sufficiently so that the trend of cooling with film thickness is dominant and thinner films are colder. Note that this non-monotonic behavior depends on the relative strengths of the heat source and the cooling term and may change if different forms are used to describe these effects. The explanation given above is the primary reason for the behavior seen in Fig.~\ref{fig:flat_film_temp_contours}(b).

\bibliography{films}
\bibliographystyle{jfm}

\end{document}